\documentclass[twocolumn]{aastex63}

    \usepackage{array}
    \usepackage{float}
    \usepackage{xcolor,colortbl}
    \definecolor{c1}{rgb}{.96, .96, 0.96}
   \newcommand{\head}[2]{\multicolumn{1}{>{\centering\let\newline\arraybackslash}p{#1}}{{#2}}}

\graphicspath{{./}{figures/}}
\shorttitle{Cloudy Hot Jupiters}
\shortauthors{Roman et al.}

\begin{document}

\title{Clouds in Three-Dimensional Models of Hot Jupiters Over a Wide Range of Temperatures I: Thermal Structures and Broadband Phase Curve Predictions}

\correspondingauthor{Michael T.\ Roman}
\email{mr359@le.ac.uk}

\author[0000-0001-8206-2165]{Michael T.\ Roman}
\affiliation{School of Physics and Astronomy, University of Leicester, University Road, Leicester LE1 7RH, UK}

\author[0000-0002-1337-9051]{Eliza M.-R.\ Kempton}
\affil{Department of Astronomy, University of Maryland, College Park, MD 20742, USA}

\author[0000-0003-3963-9672]{Emily Rauscher}
\affiliation{Department of Astronomy, University of Michigan, 1085 South University Avenue, Ann Arbor, MI 48109, USA}

\author[0000-0001-5737-1687]{Caleb K. Harada}
\altaffiliation{NSF Graduate Research Fellow}
\affil{Department of Astronomy, University of California Berkeley, University Drive, Berkeley, CA 94720, USA}
\affil{Department of Astronomy, University of Maryland, College Park, MD 20742, USA}

\author[0000-0003-4733-6532]{Jacob L.\ Bean}
\affil{Department of Astronomy \& Astrophysics, University of Chicago, 5640 South Ellis Avenue, Chicago, IL 60637, USA}

\author[0000-0002-7352-7941]{Kevin B.\ Stevenson}
\affiliation{Johns Hopkins APL, 11100 Johns Hopkins Rd, Laurel, MD 20723, USA}

\begin{abstract}
Using a general circulation model (GCM), we investigate trends in simulated hot Jupiter atmospheres for a range of irradiation temperatures (1,500 -- 4,000 K), surface gravities (10 and 40 m s$^{-2}$), and cloud conditions. Our models include simplified temperature-dependent clouds with radiative feedback and show how different cloud compositions, vertical thicknesses, and opacities shape hot Jupiters atmospheres by potentially increasing planetary albedos, decreasing photospheric pressures and nightside temperatures, and in some cases producing strong dayside thermal inversions. With decreasing irradiation, clouds progressively form on the nightside and cooler western limb followed by the eastern limb and central dayside. We find that clouds significantly modify the radiative transport and affect the observable properties of planets colder than $T_{\mathrm{irr}} \approx$ 3,000~K ($T_{\mathrm{eq}}~\approx$~2,100 K) depending on the clouds' vertical extent. The precise strength of expected effects depends on the assumed parameters, but trends in predicted phase curves emerge from an ensemble of simulations. Clouds lead to larger phase curve amplitudes and smaller phase curve offsets at IR wavelengths, compared to cloud-free models. At optical wavelengths, we predict mostly westward phase curve offsets at intermediate temperatures ($T_{\mathrm{irr}} \approx$ 2,000 -- 3,500 K) with clouds confined to the nightside and western limb. If clouds are vertically compact (i.e.\ on order of a pressure scale height in thickness), their distributions and effects become more complicated as different condensates form at different heights --- some too deep to significantly affect the observable atmosphere. Our results have implications for interpreting the diversity of phase curve observations of planets with $T_{\mathrm{irr}} \lesssim$~3,000~K.
\end{abstract}

\keywords{hydrodynamics – infrared: planetary systems – planets and satellites: atmospheres – planets and
satellites: gaseous planets – radiative transfer – scattering}

\section{Introduction} \label{sec:intro}

\subsection{Current State of Hot Jupiter GCMs}

Three-dimensional (3-D) general circulation models (GCMs) are essential tools for predicting the thermal emission and reflected light from hot Jupiters.  While the physical treatments vary from model to model, the basic predictions from different hot Jupiter GCMs are consistent.  A super-rotating equatorial jet is established at thermal photospheric pressures, which promotes heat transport from the hot dayside to the non-irradiated nightside of the tidally locked planet and shifts the planet's hot spot (its location of maximum photospheric temperature) to the east (downwind) of the substellar point \citep[e.g.][]{Showman&Guillot2002, langton07, dobbsdixon08, rauscher+2010, Heng2011}.

Population-level trends that are expected for hot Jupiter circulation as a function of irradiation were worked out analytically by \citet{KomacekShowman2016}.  Namely, day-night temperature contrasts are expected to increase as a function of increasing planetary irradiation, while hot-spot offsets away from the substellar point are expected to decrease.  The physical mechanism for these trends, as identified by \citet{PerezBeckerShowman2013}, is that radiative timescales decrease relative to the timescale for gravity waves to propagate horizontally over planetary scales, as atmospheric temperatures increase.  Therefore, the most highly irradiated planets have relatively less opportunity to transfer heat away from the substellar point and around to the nightside of the planet before the planet re-radiates the incident stellar energy to space.  These trends are manifest in hot Jupiter GCM simulations that include the radiative energy deposition from the planet's host star as the primary physical mechanism driving atmospheric circulation \citep[e.g.][]{perna12, PerezBeckerShowman2013, showman15, kataria+2016}.  

The radiative vs.\ wave propagation timescales framework is attractive because of its simplicity and because it makes straightforward predictions that can be tested by observations. However, it has been recognized that various second-order effects, beyond simple radiative forcing, may fundamentally alter the predictions of standard hot Jupiter GCMs.  For example, planets with irradiation temperatures ($T_{\mathrm{irr}}$)\footnote{The irradiation temperature is defined as $T_{\mathrm{irr}} \equiv T_{\mathrm{\star}} \sqrt{R_{\star}/a}$, where $T_{\mathrm{\star}}$ is the host star effective temperature, $R_{\star}$ is the host star radius, and $a$ is the planet's orbital semi-major axis. It is related to the zero albedo, full planet redistribution equilibrium temperature, $T_{\mathrm{eq}} = T_{\mathrm{irr}} / \sqrt{2}$.} in excess of $\sim$2,100 K could have altered thermal structures and global-scale hydrodynamic flows due to interaction between a partially ionized atmosphere and the planet's magnetic field.  The impact of such magnetic interactions on atmospheric circulation is generally predicted to reduce windspeeds and therefore inhibit day-night heat exchange \citep{Perna2010, Rogers2014}.  Additionally, for the most highly irradiated hot Jupiters with $T_{\mathrm{irr}} \gtrsim 3,500$~K (the so-called ultra-hot Jupiters), H$_2$ can dissociate on the planets' daysides and subsequently recombine on the nightsides.  Since dissociation requires energy, and recombination releases energy, these processes cool the dayside and heat the nightside, producing a more homogenized planetary temperature on global scales \citep{bell18, komacek18, tan19, mansfield20}. Finally, at lower levels of planetary irradiation, aerosols in the form of clouds and perhaps hazes can form, impacting atmospheric dynamics, thermal structures, and phase curve observables.  This is the subject of our current work, in which we investigate the effects of clouds on GCM simulations and observable properties of hot Jupiter atmospheres.

\subsection{Clouds in Hot Jupiter GCMs}

There have been multiple efforts recently to account for aerosol formation in hot Jupiter GCMs. A hierarchy of modeling approaches has emerged, ranging from simpler idealized parameterizations of presumed clouds \citep[e.g.][]{Kataria2015,Oreshenko2015, parmentier+2016, Roman&Rauscher2017}, to complex modeling of detailed cloud chemistry and microphysics \citep[e.g.][]{lee15, lee+2016, lines+2018}. 

On the simpler end of this spectrum, investigators have parameterized clouds as scatterers with spatial distributions that are either prescribed \citep[e.g.][]{Roman&Rauscher2017} or roughly determined based on expectations of thermochemical equilibrium. In practice, this typically involves comparing modeled atmospheric temperature fields to expected condensation/deposition curves for likely cloud species given an assumed chemistry. In many cases, the modeled temperature field in which the cloud distributions are evaluated is taken from a completed GCM simulation of a clear atmosphere --- a technique referred to as \textit{post-processed} cloud modeling \citep{kataria+2016,oreshenko+2016,parmentier+2016}.  In contrast, others \citep[e.g.][]{Roman&Rauscher2019} have evaluated condensation curves against the temperature field throughout their simulations, actively processing the cloud distribution and taking into account the effects clouds have in shaping the temperatures \emph{during} the evolution of the GCM. In doing so, \cite{Roman&Rauscher2019} found that this radiative feedback can have substantial effects on hot Jupiter thermal structures, especially in the case of thicker clouds, consistent with findings of \citet{lines+2018}.

On the opposite end of the modeling spectrum, more extensive cloud microphysics calculations have been incorporated into a handful of hot Jupiter GCMs \citep[e.g.][]{lee15, lee+2016, lines+2018, helling19}.  These models directly describe and track a variety of physical processes such as cloud nucleation, growth, evaporation, and rainout. Crucially, these detailed models help to reveal the relative importance and overall effect of various physical processes on the resulting cloud properties and distributions.  

In general, the more self-consistent cloud models are vastly more computationally expensive, so they have only been run for a small number of individual planetary realizations over shorter timescales.  Such models can also result in a loss of intuitive grasp for which processes are driving key trends.  Our current study aims to model a large number of hot Jupiters spanning a range in $T_{\mathrm{irr}}$ and surface gravity ($g$), so it necessitates a parameterized cloud treatment.  \citet{parmentier+2016} first studied the effects of a variety of equilibrium condensates on hot Jupiter GCMs and optical phase curves as a function of $T_{\mathrm{irr}}$, but without the inclusion of radiative feedback.  In the present work, we examine  effects of clouds with varying assumptions over a wider range of temperatures and conditions using a model that includes radiative feedback.

\subsection{Evidence for Clouds Impacting Hot Jupiter Phase Curves}

There is circumstantial evidence that clouds are a driving force shaping hot Jupiter phase curve observations.  At temperatures below $\sim$2,000 K various cloud species can begin to condense \citep[e.g.][]{Mbarek&Kempton2016}, and it is therefore expected that even for hotter planets, clouds will form on their cooler nightsides.  This prediction was first seen to potentially play out in the spectral phase curve of the hot Jupiter WASP-43b, which showed less nightside flux at IR wavelengths than would otherwise be expected for a planet with no clouds \citep{stevenson+2014}.  This diminished nightside flux has notionally been attributed to clouds raising the photospheric altitude in that region and therefore reducing the nightside brightness temperature \citep{Kataria2015, stevenson+2017}.  The full ensemble of IR hot Jupiter phase curves to date appear to display similar behavior, with near-uniform nightside temperatures inferred \citep{keating+2019, beatty19}. Such results are not predicted by cloud-free GCMs, but could be a natural outcome of cloud formation.  

In the optical, four hot Jupiters, all notably with $T_{\mathrm{irr}} <$ 2,500 K, show \emph{westward} offsets of their peak brightness \citep{demory+2013, esteves+2015}, which is unusual given that hot Jupiter temperatures are expected to peak to the east of the substellar point.  A westward offset might be expected if reflective clouds spilled over from the nightside to the dayside on the cooler western terminator, increasing the scattered light contribution in that region, while the hotter eastern terminator remained too hot for clouds to form. Detailed modeling has shown that such asymmetric phase curves would indeed require aerosols to form exceptionally thick and reflective clouds at very high altitudes centered on the western limb in order to match the observations \citep{munoz+2015}.

Prior to the end of the \textit{Spitzer} mission, IR phase curves were obtained for a total of 31 giant planets\footnote{\url{https://irsa.ipac.caltech.edu/data/SPITZER/docs/files/spitzer/extrasolarplanets.txt}}.  Approximately half of these observations have been reported in the peer-reviewed literature to date. A smaller number of optical phase curves have additionally been reported from \textit{Kepler} and \textit{K2}, and there are a growing number from the ongoing \textit{TESS} mission.  Some trends appear to emerge --- e.g.\ peak offsets at IR wavelengths correlate with irradiation temperature \citep{zhang18}, and westward optical phase curve offsets are obtained primarily for cooler planets \citep{esteves+2015}.  At present, such trends should be interpreted with caution though because they are likely impacted by a combination of small number statistics and data systematics \citep[i.e.\ different \textit{Spitzer} data reduction pipelines can sometimes yield inconsistent phase curve amplitudes and offsets;][]{stevenson+2017,Mendoca2018,keating2020,Bell20}.  Furthermore, there is substantial scatter in the observed trends, and the trends themselves do not readily match the classic predictions of hot Jupiter GCMs (e.g. decreasing day-night contrast with decreasing $T_{\mathrm{irr}}$). It therefore currently remains challenging to disambiguate which physical mechanisms are dominating the hot Jupiter atmospheres for which phase curves are available.  

\subsection{This Study}

In this paper, we focus on the role that condensate clouds play in shaping the 3-D atmospheric structure and phase curve predictions for hot Jupiter atmospheres. In Section \ref{sec:methods}, we discuss our methods, in which we extend the work of \cite{Roman&Rauscher2019} and, following \cite{Parmentier2016}, produce a new grid of models to investigate the effects of clouds over a range of temperatures---carefully chosen here to span the set of hot Jupiters that have had phase curve observations with NASA's \textit{Spitzer Space Telescope}.  We build upon previous investigations by exploring the effects of several different implementations of our cloud models, testing sensitivity to assumptions regarding chemistry, surface gravity, and vertical mixing as a function of irradiation temperatures, while including the important effects of radiative feedback from multiple cloud-forming species \citep{Roman&Rauscher2019,Lines+2019}. In Section~\ref{sec:Results}, we present and discuss the resulting suite of 93 simulations, which collectively attempt to encompass the observations of hot Jupiter atmospheres that are impacted by clouds. By performing an intercomparison of our results for a range of cloudy and clear atmospheres, we reveal how clouds can alter the thermal structures, atmospheric dynamics, and observable properties of variously irradiated hot Jupiter atmospheres.  Finally, in Section \ref{sec:discussion}, we summarize our findings and conclude with a discussion of modelling limitations, observational implications, and plans for future work.

\section{Methods}\label{sec:methods}

To simulate atmospheric temperatures, winds, radiative fluxes, and idealized cloud distributions, we employ a hot Jupiter GCM  \citep{rauscher+2010,rauscher+2012,rauscher13}, including more recent updates to account for the radiative feedback and scattering from clouds \citep{Roman&Rauscher2017,Roman&Rauscher2019}. 
The GCM's dynamical core solves the primitive equations of meteorology and is coupled to a double-gray, two-stream radiative transfer scheme based on \citet{toon+1989}.  Gaseous opacities are chosen for the visible and infrared channels following \cite{Roman&Rauscher2017,Roman&Rauscher2019}, with gaseous Rayleigh scattering included in the visible. Simple 1-D analytical solutions to a globally averaged temperature profile assuming these same double-grey gaseous opacities are used to initialize the model for each combination of irradiation temperature and surface gravity \citep[][]{guillot2010radiative}. Relevant modeling parameters are provided in Tables \ref{table:modparams} and \ref{table:modparams2}.

\begin{deluxetable*}{lccc}[th!]
\centering
\tabletypesize{\footnotesize}
\tablecolumns{4} 
\tablewidth{4in}
\tablecaption{ Fixed Model Parameters}
 \label{table:modparams}
\tablehead{
 \colhead{Parameter} & \colhead{Value} & \colhead{Units} & \colhead{Comment} 
}
 \startdata 
\it{    Orbital/Dynamical}\\
Radius of the planet, $R_p$ & $9.65\times 10^7$ & m & 1.35 $R_{jup}$ \\
Gravitational acceleration, $g$ & 10 $\&$ 40 & m s$^{-2}$ &  Two simulations for each $T_{\mathrm{irr}}$ \\
Rotation rate, $\Omega$ & $3.85 \times 10^{-5}$ & s$^{-1}$ & tidally synchronized, 1.89 day orbit\\\\
\it{    Radiative Transfer}\\
Specific gas constant, $\mathcal{R}$ & 3523 & J kg$^{-1}$ K$^{-1}$ & assumed $H_2$-rich\\
Ratio of gas constant to heat capacity, $\mathcal{R}/c_P$ & 0.286 & -- & assumed diatomic \\
Internal heat flux, $F_{\uparrow \mathrm{IR}, \mathrm{int}}$ & 3544 & W m$^{-2}$  & from modeled 1-D T-profile\\
Gaseous visible absorption coefficient, $\kappa_{\mathrm{vis}}$  & $1.57 \times 10^{-3}$  & cm$^2$ g$^{-1}$ & constant, from modeled 1-D T-profile \\
Gaseous visible scattering coefficient, $\kappa_{\mathrm{Ray}}$  & $8.09 \times 10^{-4}$  & cm$^2$ g$^{-1}$ & Rayleigh scattering for a spherical albedo of $\sim$0.15 \\
Gaseous infrared absorption coefficient, $\kappa_{\mathrm{IR}}$  & $1.08 \times 10^{-2}$  & cm$^2$ g$^{-1}$  & constant, from modeled 1-D T-profile\\\\
\it{    Model Resolution}\\
Vertical layers & 50 & --\\
Bottom of modeling domain pressure &  $\sim$100 & bar \\
Top of modeling domain pressure & 5.7 $\times$ $10^{-5}$ & bar\\
Horizontal Resolution & T31 & -- & corresponds to  $\sim$48 lat $\times$  $\sim$96 lon\\
Dynamical Temporal Resolution & 4800 & time steps/day & \\
Radiative Transfer Temporal Resolution  & 1200 & time steps/day & heating rates updated every 4 timesteps\\
Simulated Time & 2000 & planet days & 2000 revolutions\\
\enddata
% \vspace{-0.8cm}
%\label{table:modparams}
\end{deluxetable*}

%TABLE OF GRID PARAMETERS 
%%%version with columns for 100% mass
\begin{deluxetable*}{cccccccccc}[th!]
\centering
\tabletypesize{\footnotesize}
\tablecolumns{10} 
\tablewidth{4.in}
\tablecaption{ Variable Model Parameters and Suite Overview}
\label{table:gridparams}
\tablehead{ 
 \head{2.6cm}{\vspace{-0.1cm} Irradiation  [Equilibrium] Temperature (K)} & 
 \head{2.cm}{\vspace{0.1cm} Incident flux (W m$^{-2})$} & 
 \head{1.2cm}{\vspace{0.1cm} Gravity (m s$^{-2}$)} &
 \head{0.8cm}{\vspace{0.1cm} Clear?} &  
 \head{1.2cm}{\vspace{0.1cm} Extended?} & 
 \head{1.2cm}{\vspace{0.1cm} Compact?} & 
 \head{1.5cm}{\vspace{-0.1cm} Extended Nucleation Limited?} &
 \head{1.5cm}{\vspace{-0.1cm}Compact Nucleation Limited?} &
 \head{1.2cm}{\vspace{-0.1cm}Extended, 100\% mass?} &
 \head{1.2cm}{\vspace{-0.1cm}Compact, 100\% mass?} 
 }
\startdata 
 1500  [1061] & $2.87062\times10^5$ & 10 &\checkmark & \checkmark & \checkmark & \checkmark & \checkmark & \checkmark & \checkmark \\ \vspace{.2cm}
  \ &  & 40    & \checkmark & \checkmark & \checkmark &  &  & \\
\ 1750 [1237]  &  $5.31818\times10^5$ & 10 & \checkmark & \checkmark & \checkmark  & \checkmark & \checkmark & \checkmark & \checkmark \\\vspace{.2cm}
\  & & {40}   & {\checkmark} & {\checkmark} &  &  \\
\ 2000 [1414] & $9.07259\times10^5$ & 10 & \checkmark & \checkmark & \checkmark & \checkmark & \checkmark & \checkmark & \checkmark\\\vspace{.2cm}
\  & & 40    & \checkmark & \checkmark & \checkmark &  \\
\ 2250 [1591] & $1.45325\times10^6$ & 10 & \checkmark & \checkmark & \checkmark & \checkmark & \checkmark & \checkmark & \checkmark  \\\vspace{.2cm}
\  & &  40   & \checkmark & \checkmark &  &  \\
\ 2500 [1768]  &$2.21499\times10^6$ & 10  & \checkmark & \checkmark & \checkmark & \checkmark & \checkmark & \checkmark & \checkmark \\\vspace{.2cm}
\  & & 40   & \checkmark & \checkmark & \checkmark &  \\
\ 2750 [1945] & $3.24296\times10^6$ & 10 & \checkmark & \checkmark &  &  &  & \checkmark  \\\vspace{.2cm}
\  & & 40    & \checkmark & \checkmark   \\
\ 3000 [2121] & $4.59300\times10^6$ & 10 & \checkmark & \checkmark & \checkmark & \checkmark & \checkmark & \checkmark & \checkmark \\\vspace{.2cm}
\ & & 40  &\checkmark & \checkmark & \checkmark   \\
\ 3250 [2298] & $6.32622\times10^6$ & 10 & \checkmark & \checkmark & & &  & \checkmark    \\\vspace{.2cm}
\  & & 40  & \checkmark & \checkmark   \\
\ 3500 [2475]  &$8.50909\times10^6$ & 10  & \checkmark & \checkmark & \checkmark & \checkmark & \checkmark & \checkmark & \checkmark \\\vspace{.2cm}
\  & & 40   & \checkmark & \checkmark & \checkmark & \\
\ 3750 [2652] & $1.12134\times10^7$ & 10 & \checkmark & \checkmark & & & & \checkmark    \\\vspace{.2cm}
\  & & 40    & \checkmark & \checkmark &  &  \\
\ 4000 [2828]  & $1.45161\times10^7$ & 10 & \checkmark & \checkmark & \checkmark & \checkmark & \checkmark & \checkmark & \checkmark \\\vspace{.2cm}
\ & & 40   & \checkmark & \checkmark &  \checkmark \\
 \enddata
% \vspace{-0.8cm}
\tablecomments{Equilibrium temperatures are approximate and assume zero albedo and full heat redistribution. Incident fluxes are given at the substellar point. Check marks indicate which of the following cases were simulated: clear, extended cloud, compact cloud, extended nucleation-limited cloud, compact nucleation-limited cloud, extended 100\% condensed cloud, or compact 100\% condensed cloud.}

 \label{table:modparams2}
\end{deluxetable*}

\subsection{Cloud Modeling}\label{sec:cloudmodel}
Following \cite{Roman&Rauscher2019}, clouds are modeled as temperature-dependent sources of optical thickness with prescribed scattering properties appropriate for potential condensates. Starting with a clear atmosphere, clouds are formed when and where the simulated atmospheric temperature at a given pressure falls below a given compound's assumed phase transition temperature, which we will simply refer to as the condensation temperature (irrespective of whether the gas forms  solid or liquid particles). If the atmospheric temperature subsequently rises above the condensation temperature, clouds are removed. This criterion is evaluated locally at each time step of the model, allowing clouds to form or dissipate as the temperatures evolve in time. Importantly, clouds themselves actively influence the temperature field by means of scattering and absorption of both visible and thermal radiation, thus permitting feedback between the temperature field and cloud distribution \citep{lines+2018,Roman&Rauscher2019,Lines+2019}.

Thirteen different potential cloud species are included in our models (see Table \ref{table:clouds} and Appendix \ref{sec:appendix}), covering a range of condensation temperatures and scattering properties. Simplified condensation curves are obtained from \cite{Mbarek&Kempton2016} for a solar composition atmosphere in chemical equilibrium with fully efficient rainout of condensibles. For simplicity, we assume that the gaseous mole fractions are constant and uniform throughout the atmosphere regardless of whether clouds form or not, insofar as they affect the gaseous opacities, specific gas constant, and spatial variations of clouds. Likewise, potential complications of cloud chemistry, such as the temperature dependence of the preferred oxide (e.g.\ CaTiO$_3$ vs.\ Ca$_4$Ti$_3$O$_{10}$) and interactions between different condensates, are ignored. 

When temperatures fall significantly below the condensation temperature, the mass of condensate formed is assumed to equal a proportion of the total gas mass per unit area at the given pressure level $p$, evaluated as  

\begin{equation} \label{eq1}
{m_c(p)}=\frac{\Delta P(p)}{g} {\chi_g} \frac{\mu_g}{\overline{\mu}}f
\end{equation}
where $m_c$ is the mass of condensate per unit area, ${\Delta P}$ is the change in pressure across the layer, ${\chi_g}$ is the mole fraction of the relevant gas species, ${\mu_g}$ is the molecular weight of the condensed molecular compound, ${\overline{\mu}}$ is the mean molecular weight of the hydrogen-rich atmosphere (assumed to be 2.36 g/mol), and $f$ is the fraction that condenses. The fraction of gas mass that ultimately forms and persists as a cloud will depend on several microphysical processes (including nucleation, coalescence, entrainment, and fallout \citep[e.g.][]{ Rossow1978,lunine1989effect,ackerman2001precipitating,Marley2013exoclouds,gao2020aerosol} for which observational and theoretical constraints are lacking; therefore, we consider two different values of the condensing fraction $f$: 10$\%$ and 100$\%$, as discussed below. In each case, the gaseous mole fraction is simply taken to equal the mole fraction of the stoichiometrically limiting atom for each cloud species, considering relative atomic abundances and atomic ratios (see Table \ref{table:clouds}). Note that this neglects potentially large differences in particle nucleation rates \citep{gao2020aerosol}, which we address in Section \ref{sec:suite}.

The mass of each component condensate is then converted to an aerosol normal optical depth $\tau_c$ via the expression 

\begin{equation} \label{eq2}
{{\tau_c(p)}}=\frac{3 {m_c(p)} {Q_e(p)}}{4 r(p) {\rho}}
\end{equation}
where ${\rho}$ is the particle density, and ${r}$ and ${Q_e}$ are the aerosol particle radius and particle scattering extinction efficiency within the layer centered on pressure $p$. 

With this scheme as described, the cloud's vertical extent is only limited by the temperatures or boundaries of the model domain while the cloud optical depth only diminishes with height because the atmospheric pressure decreases with height (essentially equivalent to the ``frozen-in" models of \cite{lunine1989abundance}). This is a simplification that neglects any reduction in mole fraction that would result from condensing vapor, which in reality may limit the vertical extent of the condensate cloud. The degree to which saturation and other microphysical processes shape the cloud vertical extent strongly depends on the amount of vertical mixing in the atmosphere, which remains highly uncertain with estimates of eddy diffusivity varying by orders of magnitude \citep{Moses2013,Parmentier2013, Agundez2014}. Yet, as \cite{Roman&Rauscher2019} showed, whether clouds are vertically extended or compact can significantly affect the observed phase curves.  Considering this, we also allow modeling of more vertically compact clouds for comparison.  In these compact cases, the opacity within the base layers of the cloud is computed as above, but the condensate optical depth is truncated to zero at one pressure scale height above the cloud's base. Simulations using each cloud assumption are performed as discussed in detail in Section \ref{sec:suite}.

Overall, this cloud modeling follows the approach taken by \cite{Roman&Rauscher2019} but with a few notable differences. Whereas \cite{Roman&Rauscher2019} included only four species of clouds (MnS, MgSiO$_3$, Fe, and Al$_2$O$_3$), we now include up to 13 different species, allowing for a richer range of cloudy effects over a wider range of temperatures.  Even with just four cloud species, \cite{Roman&Rauscher2019} limited maximum cloud masses to just 10$\%$ of the potential condensable masses after encountering numerical instabilities caused by instantaneous large changes in cloud opacities. This problem is mitigated in the present work by gradually ramping up the heating rates over the first fifty days of the simulation and introducing a more gradual transition in cloud formation and dissipation. Similarly, for improved numerical stability the cloud mass is tapered at the very top of the atmosphere such that any resulting cloud opacity exponentially diminishes across the five uppermost vertical layers in the model (corresponding to a tapering above the 0.3-mbar height). Though we still retain the 10$\%$ condensation fraction for a majority of our simulations, these modifications now allow us to also include select additional simulations in which 100$\%$ of the vapor mass is potentially condensed for comparison (see Section~\ref{sec:suite} for a summary of cases simulated). In either case, clouds only reach their full mass when temperatures fall more than 10 K below the condensation temperature. For temperature differences of less than 10 K, the cloud masses are scaled linearly as a fraction of their potential mass, providing a more gradual and numerically stable transition at temperatures where the partial pressure just marginally exceeds the saturation pressure. 

%TABLE OF CLOUDS
\begin{deluxetable*}{lccll}
\centering
\tablecolumns{5}
\tablecaption{Cloud Properties}
\tablehead{
\colhead{Cloud Compound} & \colhead{Mole Fraction}  & \colhead{Density (kg m$^{-3}$)} & \colhead{Visible Refractive Indices} & \colhead{Thermal Indices}}
\startdata
%\\
%\multicolumn{5}{c}{\emph{New}}\\
%\\
KCl & 1.23$\times$10$^{-7}$ & 1.98$\times$10$^{3}$ & (1.49) + i(7.71$\times$10$^{-11}$) &  (1.47) + i(2.53$\times$10$^{-11}$)\\
ZnS & 4.06$\times$10$^{-8}$ &  4.09$\times$10$^{3}$ & (2.35) + i(3.62$\times$10$^{-6}$) & (2.25) + i(3.37$\times$10$^{-6}$) \\
Na$_2$S & 9.35$\times$10$^{-7}$  & 1.86$\times$10$^{3}$ & (1.80) + i(1.26$\times$10$^{-2}$) & (1.74) + i(1.12$\times$10$^{-2}$) \\
MnS & 3.11$\times$10$^{-7}$ &  4.00$\times$10$^{3}$ & (2.81) + i(4.89$\times$10$^{-4}$) & (2.61) + i(1.00$\times$10$^{-5}$) \\
Cr$_2$O$_3$ & 4.40$\times$10$^{-7}$ &  5.22$\times$10$^{3}$ & (3.48) + i(4.36) & (3.91) + i(1.71$\times$10$^{1}$) \\
SiO$_2$ & 3.26$\times$10$^{-5}$ &  2.65$\times$10$^{3}$ & (1.54) + i(8.25$\times$10$^{-6}$) & (1.41) + i(1.20$\times$10$^{-3}$) \\
Mg$_2$SiO$_4$ & 1.75$\times$10$^{-5}$ &  3.27$\times$10$^{3}$ & (1.62) + i(1.30$\times$10$^{-4}$) & (1.55) + i(8.43$\times$10$^{-3}$) \\
VO & 9.56$\times$10$^{-9}$ &  5.76$\times$10$^{3}$ & (1.80) + i(7.10$\times$10$^{-1}$) & (5.38) + i(5.96) \\
Ni & 1.61$\times$10$^{-6}$ & 8.90$\times$10$^{3}$ &  (1.99) + i(4.26) & (4.46) + i(1.25$\times$10$^{1}$) \\
Fe & 2.94$\times$10$^{-5}$ &  7.90$\times$10$^{3}$ & (2.37) + i(3.36) & (4.59) + i(1.54$\times$10$^{1}$) \\
Ca$_2$SiO$_4$ & 9.95$\times$10$^{-7}$ &  3.34$\times$10$^{3}$ & (1.70) + i(9.67$\times$10$^{-4}$) & (2.13) + i(4.06$\times$10$^{-3}$) \\
CaTiO$_2$ & 7.83$\times$10$^{-8}$ &  3.98$\times$10$^{3}$ & (2.30) + i(9.67$\times$10$^{-4}$) & (2.13) + i(4.06$\times$10$^{-3}$) \\
Al$_2$O$_3$ & 1.39$\times$10$^{-6}$ &  3.95$\times$10$^{3}$ & (1.59) + i(3.62$\times$10$^{-2}$) & (1.47) + i(1.84$\times$10$^{-2}$) \\
\enddata
\tablecomments{Gaseous mole fractions ($\chi_g$) are based on the atomic abundances of \cite{burrows1999chemical} and \cite{anders1989abundances} for limiting atoms, divided by their atomic ratios in cloud compounds. Indices of refraction were taken from: \cite{al2007optical} for Cr$_2$O$_3$; \citep{wan2019optical} for VO, using values for VO$_2$; \cite{johnson1974optical} for Ni; \cite{shannon2017refractive} for Ca$_2$SiO$_4$, with lacking imaginary and thermal indices assumed equal to CaTiO$_3$; and \cite{kitzmann+2018} and references therein for the remainder.} 
\label{table:clouds}
\end{deluxetable*}
%TABLE OF CLOUDS

Additionally, we introduced a vertical gradient in cloud particles sizes.  In real atmospheres, particles tend to grow larger deeper in the atmosphere due to the the vertical gradient in pressure and kinematic viscosity \citep{roman2013saturn,Parmentier2013,lee+2017,powell+2018,lines+2018}. We use an exponential gradient that is loosely based on the trends found in the GCM simulations of \cite{lines+2018}. Particles that form in the top layers of the atmosphere are fixed at 0.1 $\mu$m, and assumed to increase exponentially in radius at pressures greater than 10 mbar, reaching nearly 80 $\mu$m in radius at the $\sim$ 100-bar base of the atmospheric domain. This vertical gradient in particle size results in a pressure dependence in the cloud particle scattering properties --- specifically the extinction efficiency, single scattering albedo, and asymmetry parameter. These scattering parameters are pre-computed using Mie theory with the particle radii and refractive indices as inputs for each compound at each pressure level \citep{de-rooij+1984, mishchenko+1999}, assuming a log-normal particle size distribution with a variance of 0.1 $\mu$m. Although detailed theoretical modeling of exoplanet clouds have predicted more complicated particle size distributions that differ from species to species \citep{powell19}, given the modest complexity of our model and absence of observational constraints, we simply choose a single, traditional size distribution based on values found and frequently assumed in studies of terrestrial clouds \citep[e.g.][]{lopez_1977lognormal,arduini2005sensitivity} to compute scattering parameters for all clouds. In locations where multiple cloud types are simultaneously present, optical depth-weighted averages of the scattering parameters are used. More details of the scattering parameters are included in Appendix~\ref{sec:appendix}. Altogether, it is the nontrivial combination of these varied scattering parameters, the relative abundances of each condensable compound, and the condensation temperatures of each compound relative to each other and the local atmosphere, that determine the net heating rates and distributions of clouds in our models. 

Our simple double-gray, temperature-dependent scheme does not attempt to include cloud microphysics, but rather it is intended to efficiently mimic plausible cloud distributions based on physical expectations. Notably, we neglect advection of cloud particles, thermodynamics of phase changes, rainout, and other fundamental processes that can alter the optical thickness and extent of clouds. The double-gray treatment itself also introduces systematic errors in the radiative transfer through clouds, with heating rates potentially off by a factor of a few in the cloudiest regions due to the absent wavelength-dependence \citep{Roman&Rauscher2019}. Ultimately, we accept that a host of physical processes beyond the scope of this model would affect cloud masses and their potential effects, and so the fully condensed masses and vertically extended clouds presented here represent an approximation of the simple upper limit on the cloud opacities.  

\subsection{Suite of Models}\label{sec:suite}
To investigate potential trends in the observable parameter space, we generate a suite of models for hot Jupiters subject to a range of instellation and cloud modeling assumptions. In general, the model parameters are selected to span the range of planets that have \textit{Spitzer} phase curve observations, excepting KELT-9b, which is an extreme outlier in $T_{\mathrm{irr}}$ and is also unlikely to harbor any clouds \citep{gaudi2017giant}. Eleven different irradiation temperatures are used to define our grid suite, ranging from 1,500 K to 4,000 K, at intervals of 250 K. This is equivalent to planetary equilibrium temperatures ranging from 1,060 K to 2,830 K, under the assumption of zero albedo and full-planet heat redistribution. These irradiances set the boundary conditions for visible flux entering the top layer of the models, while we assume a constant internal thermal flux entering the bottom boundary from below for all cases (see Tables \ref{table:modparams} and \ref{table:modparams2}).

We first define a baseline grid consisting of a cloudy and corresponding clear simulation at each $T_{irr}$ of our temperature grid. In these baseline cloudy cases, we assume clouds are vertically extended (as described in Section~\ref{sec:cloudmodel}) and achieve 10$\%$ of the potential maximum cloud mass (i.e.\ $f=0.1$ in Equation~\ref{eq1}), following \cite{Roman&Rauscher2019}. All of these baseline simulations are repeated at two different values of surface gravity --- 10~m~s$^{-2}$ and 40 m s$^{-2}$ --- to mostly span the parameter space of \textit{Spitzer}-observed hot Jupiters. 

We supplement this baseline grid with additional simulations at select irradiation temperatures and surface gravities (as specified in Table \ref{table:modparams2}) to evaluate the potential effects of different assumptions regarding the clouds --- specifically, their vertical extent, composition, and mass. In one sub-grid, we assume that all clouds are vertically compact, truncating the cloud after a single pressure scale height in thickness above its base pressure. In contrast to our extended cloud cases, these compact cloud simulations are physically representative of an atmosphere with weak vertical mixing and consequent desiccation with increasing height that limits the clouds' vertical extent. In a second sub-grid, we investigate how results will differ if the cloud composition is affected by differences in particle nucleation rates for different species. In these cases, we simply remove the ZnS, Na$_2$S, MnS, Fe, and Ni clouds entirely, given their considerably lower nucleation rates \citep{gao2020aerosol}. Although \cite{gao2020aerosol} did not state nucleation rates for Ni, we include it in this list because it is similar to Fe in size and scattering properties (although considerably less abundant). In a third grid, we combine the properties of the previous two sub-grids and simulate clouds that are both vertically compact and nucleation-limited. Going forward, we refer to these different cloud treatments as \textit{extended}, \textit{compact}, \textit{nucleation-limited}, and \textit{compact nucleation-limited}, as detailed above. 

Most of these supplemental simulations are run at a single surface gravity of 10~m~s$^{-2}$ assuming 10$\%$ of the vapor condenses, with finer grid sampling at lower irradiation temperatures where we find variations in the cloud modeling to be most pronounced. However, considering that higher gravity may reduce the cloud scale heights, we additionally model the compact cloud case at $g=40$~m~s$^{-2}$ for a majority of the irradiation temperatures.  Finally, to determine the effects of assumed cloud opacity, we additionally model a sub-grid of both extended and compact clouds assuming 100$\%$ of the vapor condenses (i.e.\ $f=1.0$ in Equation~\ref{eq1}) and $g=10$~m~s$^{-2}$ for most irradiation temperatures, as indicated in Table \ref{table:modparams2}.  In practice, clouds with 100$\%$ condensable mass push the limits of our model's numerical stability as discussed in Section \ref{toomanyclouds}, but they nonetheless illustrate an upper limit on the potential effects of clouds in our simulations. 

Aside from the irradiation temperature and surface gravity, all remaining parameters, including the planetary radius, rotation rate, and bulk composition remained unchanged across the parameter space in order to isolate trends that result from radiatively-interacting clouds. Due to the range of stellar hosts of the observed phase curve sample, we find that the orbital periods and hence planetary rotation rates are not well correlated with $T_{\mathrm{irr}}$, motivating our choice to de-couple these two parameters in our model grid.  The 1.89 day rotation period that we select lies near the median of the \textit{Spitzer}-observed hot Jupiter sample.

We nevertheless recognize and note that some of our modeling assumptions neglect fundamental changes expected across the the domain that could additionally shape trends in the observables.  For example, differences in metallicity and equilibrium chemistry would alter cloud abundances and gaseous absorption coefficients that would in turn alter expected heating rates and energy redistribution.  We discuss the relative strengths and limitations of our model grid in more detail in Section~\ref{sec:discussion}.

\section{Results}\label{sec:Results}

The 93 individual GCMs that make up our model grid represent an extensive multi-parameter data set for systematically exploring the effects of clouds on \mbox{3-D} hot Jupiter atmospheres.  To parse the results, below, we first focus on the main effects of clouds as they behave in our baseline cloud grid, which is the vertically extended cloud implementation with 10\% condensation efficiency, by mass.  We select this as our baseline case because it produces some of the most extreme instances of radiative feedback while balancing against considerations of numerical stability.  We then explore additional dependencies of our results on the details of our cloud model by comparing between the extended, compact, nucleation-limited, and 100\% condensed cloud implementations.  Finally, we identify and diagnose observable trends in broadband phase curves that should be associated with the presence of clouds.

\subsection{Baseline Results --- Extended Clouds}

\begin{figure*}[t!]
    \centering
    \includegraphics[scale=0.47]{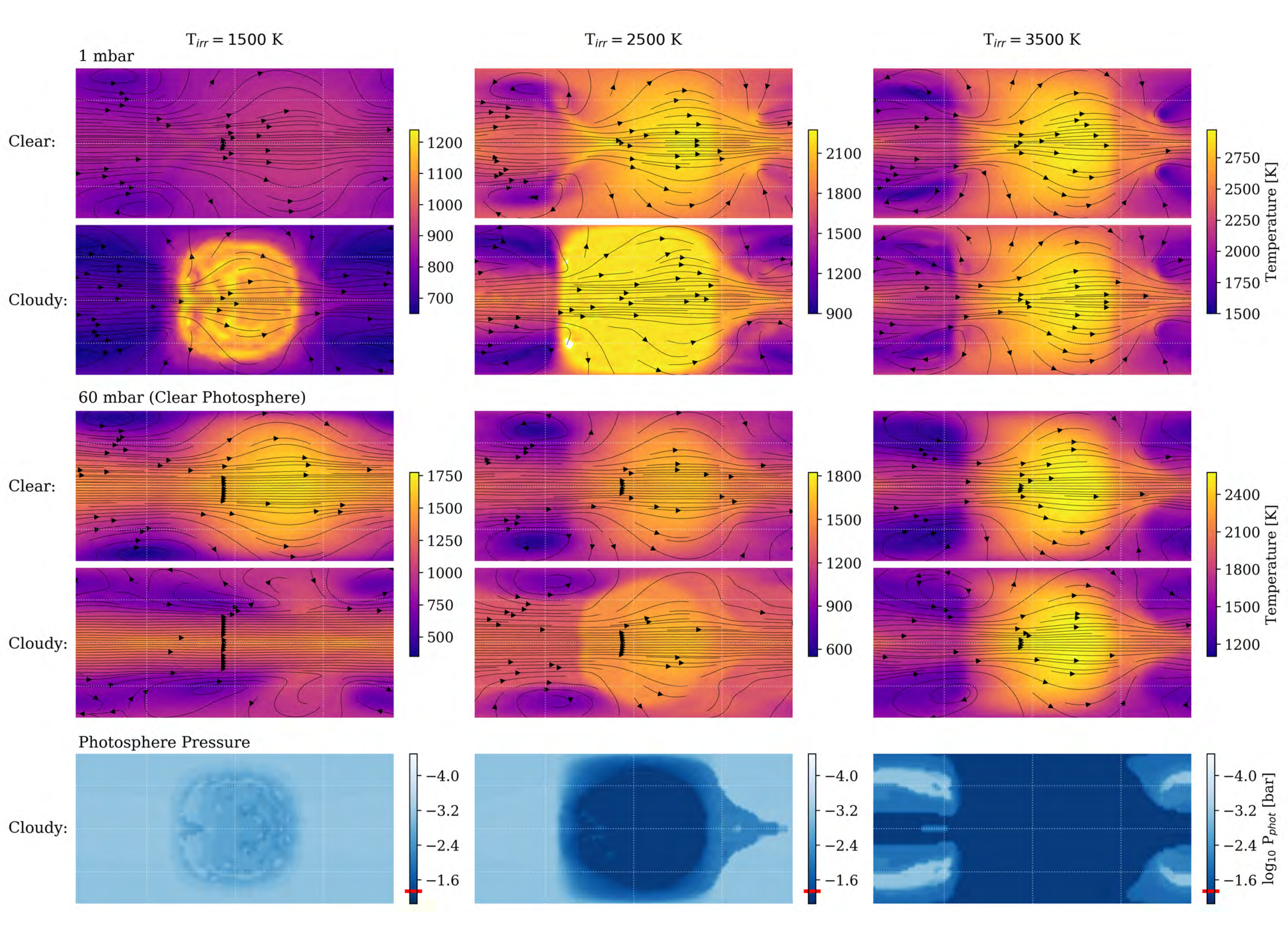}
    \caption{Temperature and IR photospheric pressure maps for our clear atmosphere and extended cloud model grids. In all panels, the substellar point is located at the center of the plot.  The $x$-axis is longitude, and the $y$-axis is latitude, with the entire global range shown. Temperature maps are shown at pressures of 1 mbar (top two rows) and 60 mbar --- roughly the pressure of the clear-atmosphere IR photosphere (3rd and 4th rows); with the clear model shown in the upper of the two panels, and the cloudy model shown immediately below, as labeled on the far left.  Note the variable temperature scale for each pair of plots. %Photospheric pressure ($P_{\mathrm{phot}}$) maps are shown for the cloudy models only (bottom row). 
    Maps indicating the pressure of the effective IR photosphere ($P_{\mathrm{phot}}$) in the cloudy models are shown below (bottom row). The clear-atmosphere photospheric pressure of 60 mbar is indicated on the color bar with a red dash. The maps of $P_{\mathrm{phot}}$ directly trace the cloud optical depths shown in Figure~\ref{fig:tinversion}. Clouds boost the photosphere to higher in the atmosphere, corresponding to lower pressures.  For these extended cloud models, the clouds cause dayside thermal inversions and upper atmosphere heating, accompanied by nightside cooling.  They furthermore homogenize the temperatures deeper in the atmosphere.  This figure shows simulations with a surface gravity of 10 m s$^{-2}$, but results are qualitatively similar for $g=40$~m~s$^{-2}$, although with greater photospheric pressures.}  
    \label{fig:temp_maps}
\end{figure*}

The 22 extended cloud models with 10\% condensation efficiency (11 values of $T_{\mathrm{irr}}$ $\times$ 2 values of $g$) make up our baseline cloudy GCM grid.  Here we compare our baseline cloudy GCM realizations to an equivalent grid of 22 clear-atmosphere models and comment on the impact of clouds.  First, our clear atmosphere models show the behavior that we have come to expect of hot Jupiter GCMs.  The flow pattern at pressures near the IR photosphere (i.e.\ where the contribution to the outgoing thermal emission is greatest) is dominated by a strong eastward equatorial jet (Figure~\ref{fig:temp_maps}, 3rd row of maps).  Large day-night temperature contrasts and small offsets of peak brightness (from the substellar point) are associated with higher values of $T_{\mathrm{irr}}$.  Higher in the atmosphere, the flow is more strongly day-to-night, where short radiative timescales regulate the atmospheric circulation (Figure~\ref{fig:temp_maps}, 1st row of maps).

The inclusion of extended clouds in our GCMs alters the behavior of the atmosphere in a number of key ways.
First of all, on a global scale, clouds increase the planet's Bond albedo by increasing scattered incident light at visible wavelengths, which in turn reduces the (net global) thermal emission in the IR as less stellar radiation is absorbed by the atmosphere (Figures~\ref{fig:cloudgrid} and \ref{fig:cloudgrid_night}).
Within the atmosphere, scattering from the clouds significantly impacts the predicted thermal structures (Figures~\ref{fig:temp_maps}-\ref{fig:tinversion}) by effectively redistributing received and emitted radiation. 

Furthermore, several of the potential condensates form cloud particles that directly absorb stellar light at optical wavelengths --- most notably the massive iron clouds and, to a lesser extent, the high-temperature corundum (Al$_2$O$_3$) clouds. This aerosol absorption directly modifies the temperature profiles by increasing the magnitude of visible heating rates on the irradiated dayside. At its most intense, aerosol heating produces sharp spikes in temperature profiles with values that tend to approach and oscillate around the absorbing cloud's condensation temperature as clouds cyclically form and vaporize. This behavior can be seen in the temperature profiles of Figure \ref{fig:t_p_cond}, with additional examples and discussion provided in Appendix \ref{sec:appendix_tp}. By increasing heating rates nearer the top of the atmosphere, aerosol heating also creates thermal inversions on the planets' daysides, as shown additionally in the thermal inversions maps in Figure~\ref{fig:tinversion}. 
Conversely, on the nightsides where no direct stellar radiation is received, clouds cool the upper atmosphere by trapping heat at depth where it is redistributed efficiently by strong zonal winds.  This leads to a more homogenized temperature structure at depth but much stronger day-night temperature contrasts higher in the atmosphere.  

Finally, wherever optically thick clouds are present, they serve to push the IR photosphere to lower pressures (higher altitudes), where the atmosphere is thinner and radiative time constants are likely shorter (Figure~\ref{fig:temp_maps}, bottom row of maps).  With our assumed gaseous opacities and a surface gravity of $g=10$~m~s$^{-2}$, our modeled clouds can reduce the IR photospheric pressure (i.e. the pressure where $\tau_{IR}$ $\approx$ 2/3 at normal emission angle) from $\sim$62 mbar to as little as 2.8 mbar. Likewise, in the visible channel, the pressure to which incident stellar light can easily penetrate can be even more greatly reduced.  While the clear atmosphere reaches an optical depth of roughly 2/3 at $\sim$280 mbar (hereafter refered to as the clear \emph{visible} photosphere) when $g=10$~m~s$^{-2}$, the presence of clouds can make the effective atmosphere optical thick at pressures of less than a millibar, significantly reducing the stellar heating at greater depths.

\subsubsection{Dependency on Irradiation Temperature}
\begin{figure*}[p]
 \centering
\includegraphics[scale=0.80]{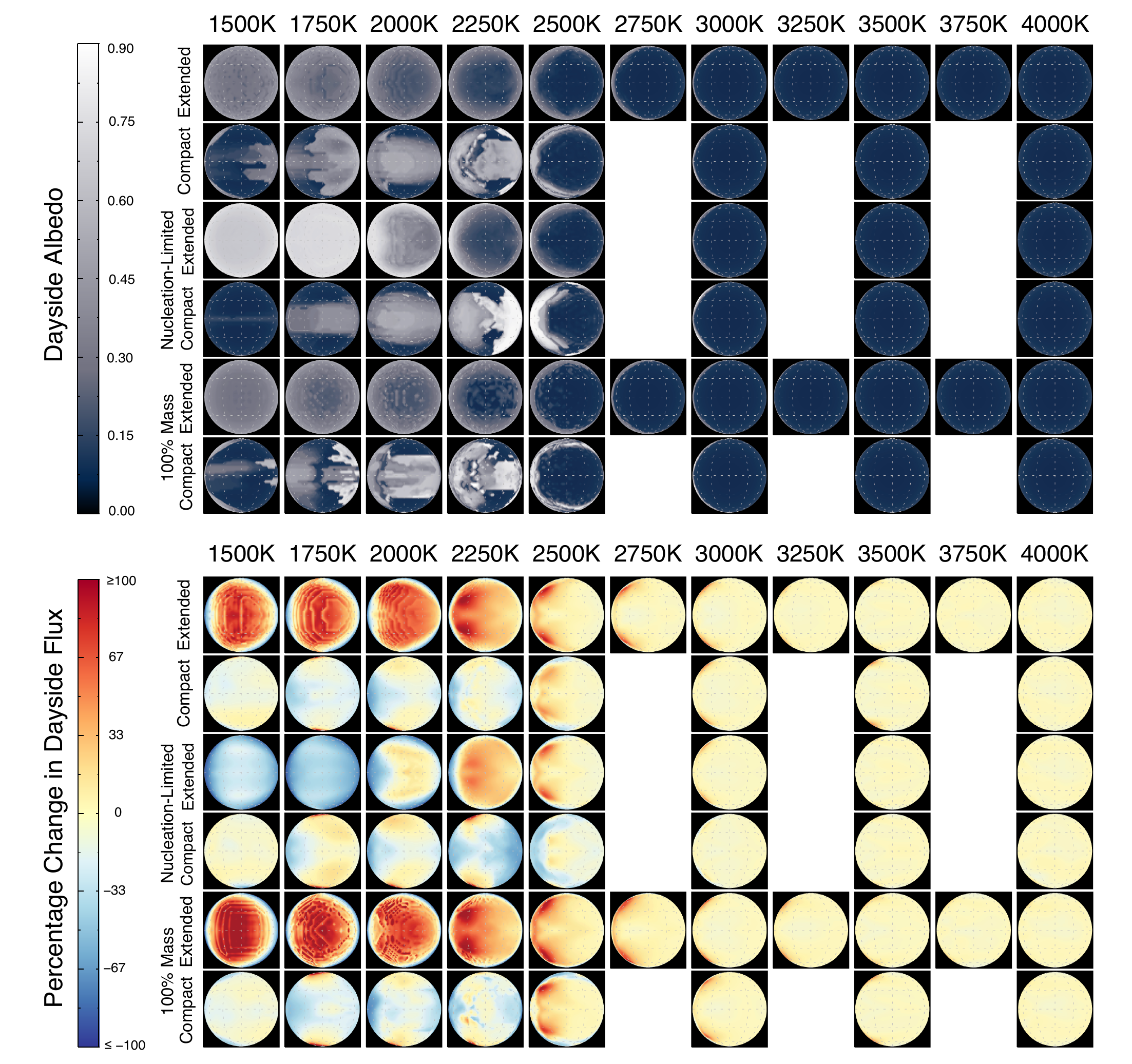}
\caption{The effects of clouds on dayside albedo (top) and thermal emission (bottom) for our different cloud models at irradiation temperatures between 1,500 K and 4,000 K. The first four rows show extended, compact, and nucleation-limited (extended and compact) cloud models assuming 10$\%$ of the vapor condenses; the following two rows show extended and compact cloud models assuming 100$\%$ condensation efficiency. Dayside albedos are calculated for each point on the disk as the ratio of upward to downward visible flux at the top of the model. For the planets with lower $T_{\mathrm{irr}}$, regional albedos are enhanced due to scattering from condensate clouds forming on the relatively cool daysides.  As $T_{\mathrm{irr}}$ increases, clouds are progressively limited in extent to the western limb and finally absent entirely from the dayside, leaving only gaseous scattering. Vertically compact clouds have different albedos as various species form over different ranges of heights. Clouds limited by nucleation rates lack absorbing Fe and have significantly greater albedos. The effect of these clouds on the dayside thermal emission is shown as a percentage change in outgoing thermal flux at the top of the atmosphere for each grid point relative to that of the corresponding clear simulations (bottom). If extended high on the dayside, our standard clouds absorb and heat the atmosphere, leading to significantly greater thermal flux in those locations. The effect is reduced when clouds are confined deeper in the atmosphere as in the compact case, and reversed when absorbing Fe and Ni clouds are removed in the nucleation-limited models. The figure shows simulations with a surface gravity of 10 m s$^{-2}$, but results are qualitatively similar for cases simulated at $g=40$ m s$^{-2}$. The corresponding nightside changes are shown in Figure \ref{fig:cloudgrid_night}.}
\label{fig:cloudgrid}
\end{figure*}

\begin{figure*}[t!]
\centering
\includegraphics[scale=0.80]{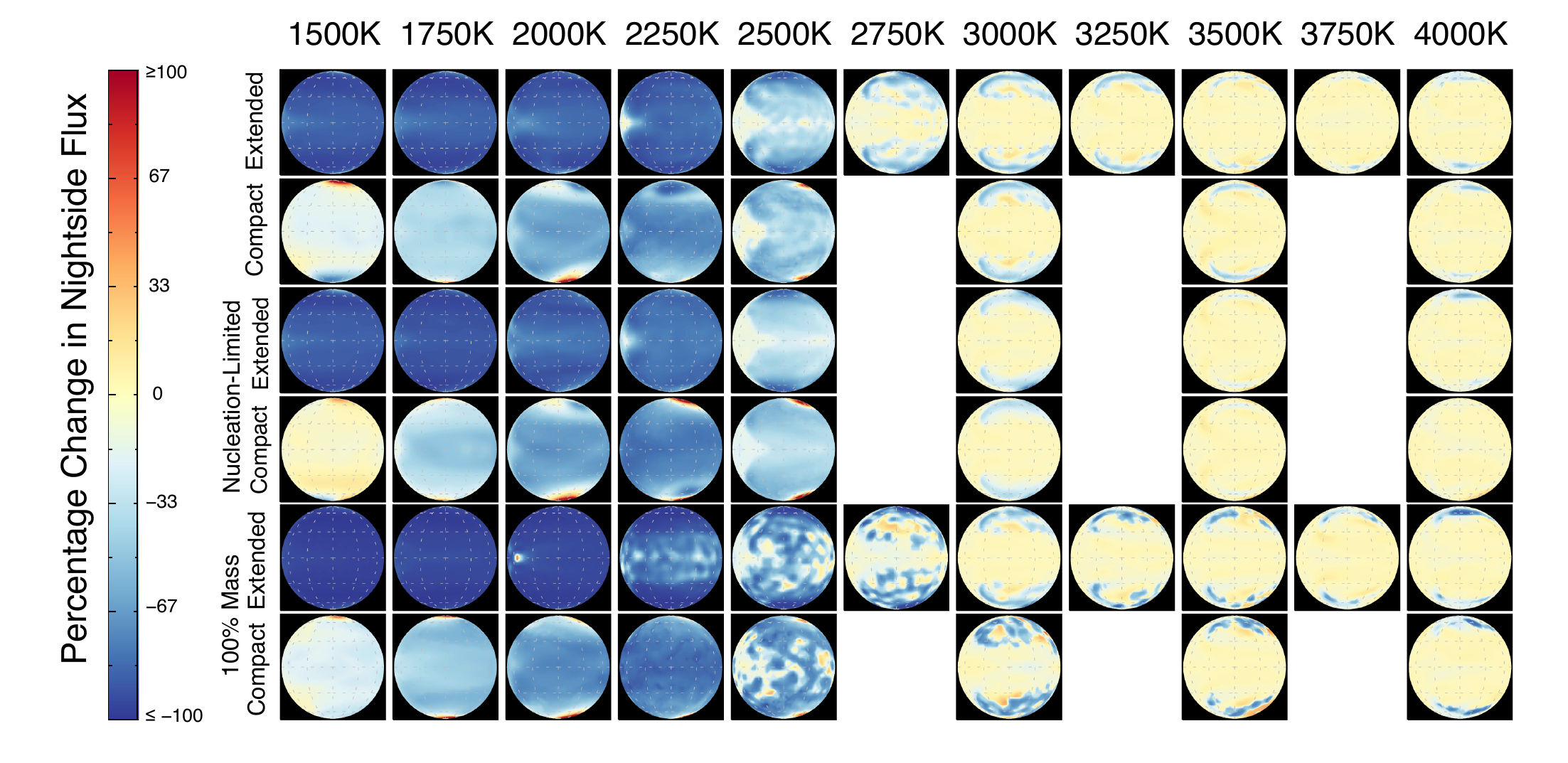}
\caption{As in Figure \ref{fig:cloudgrid}, but for the change in nightside thermal emission for our different cloud models relative to the corresponding clear models over a range of irradiation temperatures. Extended clouds at cooler temperatures significantly reduce the thermal emission from the underlying atmosphere on the nightside.  The effect is greatest when 100$\%$ of the vapor mass condenses but considerably less when clouds are vertically compact due to their lesser integrated thickness above the thermal photosphere. In the compact cases, the effects of clouds increase between 1,500 K and 2,250 K as warmer temperatures cause clouds to form at increasingly lower pressures. But for all cases, as the irradiation temperature further increases, the clouds reduce in thickness and extent starting at the warmer western equator, ultimately giving way to relatively small patches of reduced emission at the high latitudes.}
\label{fig:cloudgrid_night}
\end{figure*}

\begin{figure*}
\centering 
\includegraphics[scale=0.75, clip, trim=0in 1in 0in 1in]{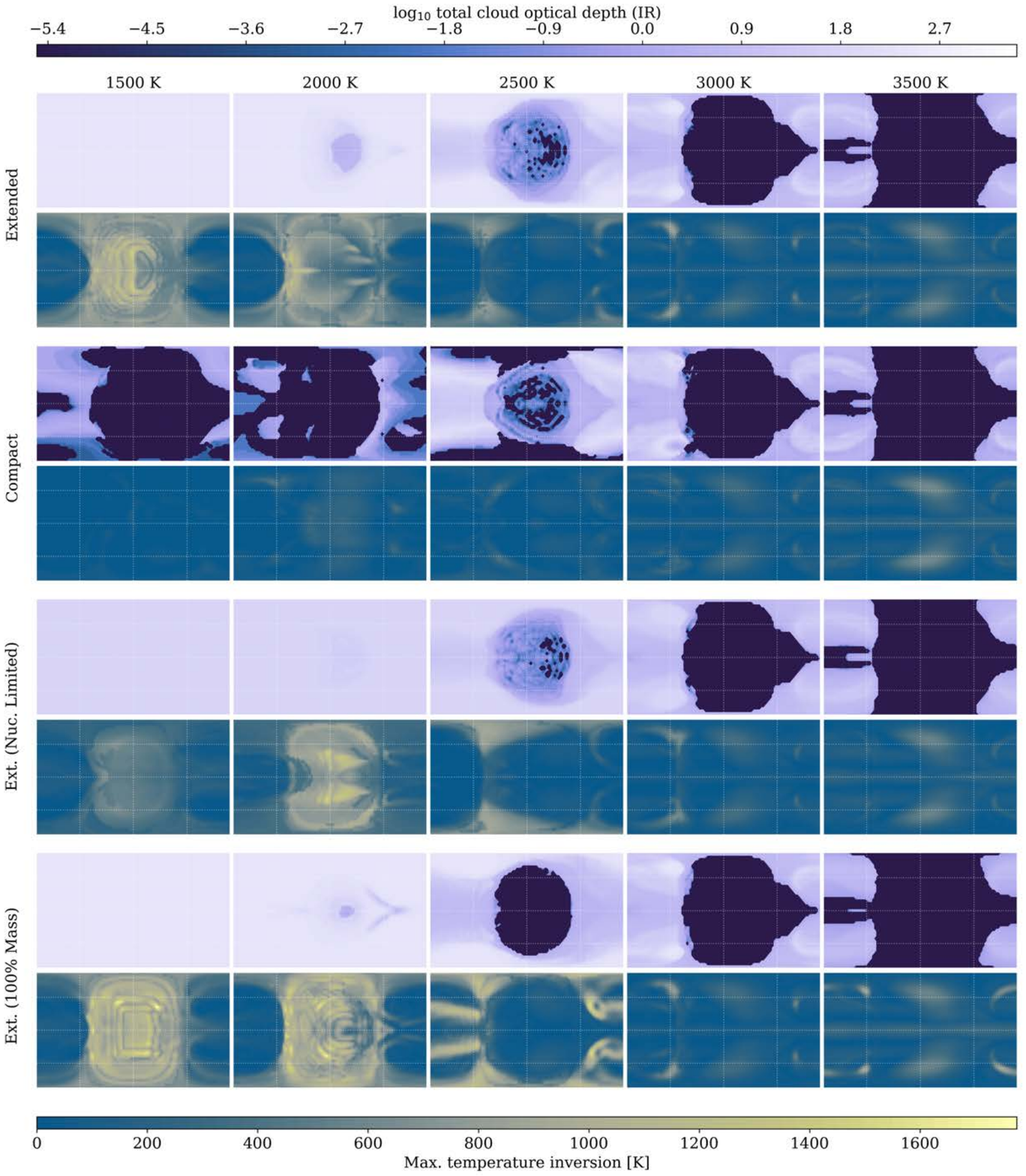}
%\gridline{\fig{tinversion_maps_g10_extended.png}{0.8\textwidth}{}}
%\gridline{\fig{tinversion_maps_g10_compact.png}{0.8\textwidth}{}}
%\gridline{\fig{tinversion_maps_g10_nuc.png}{0.8\textwidth}{}}
\caption{Maps of cloud optical depth (upper panels) and thermal inversion strength (lower panels) for five different irradiation temperatures for our extended (top), compact (second from top), nucleation-limited (third from top), and 100\% condensed (bottom) cloud models.  The integrated cloud optical depth is calculated above the cloud-free IR photosphere in each case. The thermal inversion strength is defined as in \citet{harada20} as the maximum continuous temperature increase from the bottom to the top of the atmosphere.  Strong thermal inversions in the extended cloud case correspond to the locations of thick absorptive clouds on the planets' daysides.  In the compact case, the reduced vertical extent of the clouds results in smaller cloud optical depths and less of an impact on upper-atmosphere heating, producing thermal inversions that are very weak or absent.  Thermal inversions in our nucleation-limited models are also weakened because absorptive iron clouds are not allowed to form, and remaining the dominant cloud species are relatively modest absorbers and efficient scatterers. Thermal inversions are stronger in the 100\% condensed case, as expected due to the larger column mass of cloud particles. The figure shows simulations with a surface gravity of 10 m s$^{-2}$, but results are qualitatively similar for $g=40$ m s$^{-2}$.}
\label{fig:tinversion}
\end{figure*}

The clouds themselves are not uniformly distributed, as their formation is strictly temperature dependent (see the top rows of Figures \ref{fig:cloudgrid}, \ref{fig:cloudgrid_night}, and \ref{fig:tinversion}; results for our other cloud models are deferred to Section \ref{sec:3.2}). As expected, our hottest models do not form any clouds on either their daysides and few if any on their cooler (but still quite hot) nightsides.  Planets with $T_{\mathrm{irr}} < 3,250$ K have fully clouded nightsides, and below $T_{\mathrm{irr}} = 2,250$~K, the entire planet, including its hotter dayside, becomes fully enshrouded in clouds. As clouds condense and absorb stellar radiation on the planets' daysides, the clouds heat the atmosphere, resulting in radiative feedback that causes temperatures and resulting clouds to fluctuate near the substellar regions. Consequently, in the colder models of our grid, we see patchier cloud cover and strong thermal inversions. At intermediate irradiation temperatures, the optical cloud heating increases and clouds begin to dissipate at the more highly irradiated substellar point. 
Prograde winds advect heat to the east of the irradiated substellar longitude while advecting colder gases from the planets' nightsides on the west.  This leads to clouds forming on the western dayside limb in models that are otherwise hot enough to maintain a cloud-free dayside, resulting in the crescent-like distributions seen on the daysides between 2,500 K and 3,000 K. These distributions can be seen in Figure \ref{fig:cloudgrid}, where the clouds are evident as regions of enhanced local albedo due to their back-scattering incident stellar light. 

\begin{figure*}[p]
    \centering
    \includegraphics[scale=0.35]{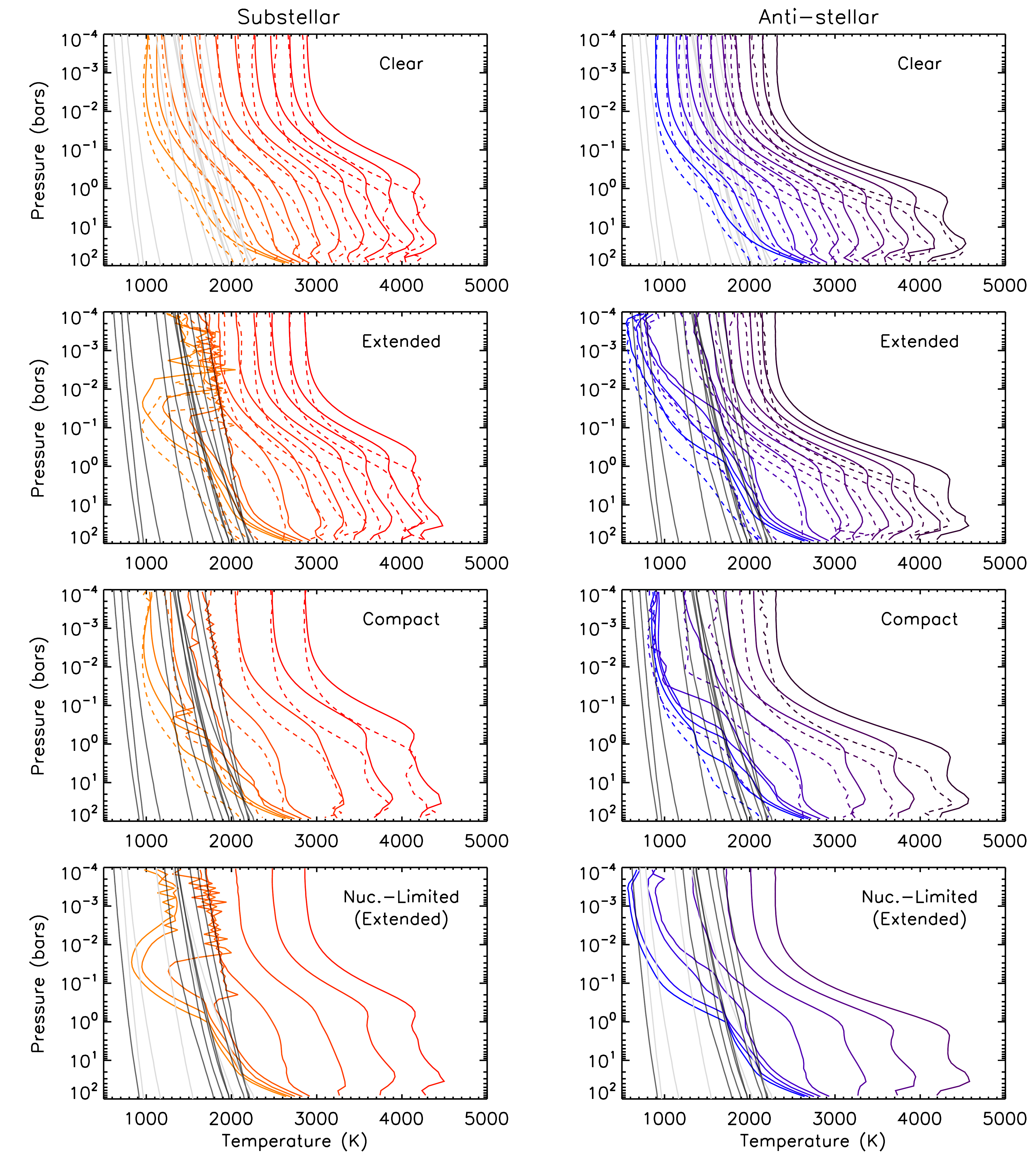}
    \caption{Vertical temperature-pressure ($T$-$P$) profiles taken from our converged GCMs at the substellar (left panels) and anti-stellar (right panels) points on the planet.  From top to bottom, results are shown for simulations with clear atmospheres, vertically extended clouds, compact clouds, and nucleation-limited (extended) clouds.  In all panels, solid lines are for cases with $g = 10$ m s$^{-2}$, and dashed lines are for cases with $g = 40$ m s$^{-2}$.  Solid dark gray lines are the condensation curves of the 13 condensible species included in our models.  From left to right along the bottom x-axis, they are KCl, ZnS, Na$_2$S, MnS, Cr$_2$O$_3$, SiO$_2$, VO, Mg$_2$SiO$_4$, Ni, CaTiO$_3$, Ca$_2$SiO$_4$, Al$_2$O$_3$, and Fe.  The same condensation curves are shown in light gray for simulations in which clouds were not included (top panels) or certain cloud species were removed from the calculation (bottom panels).  In our cloudy models, the cloud base occurs where the $T$-$P$ profile crosses a condensation curve from right to left (higher $T$ to lower~$T$), as pressure decreases.  Conversely, clouds evaporate where $T$-$P$ profiles cross in the direction of increasing temperature. Qualitatively, depending on their optical properties, dayside clouds promote thermal inversions of differing magnitudes.  On the nightside, clouds have a cooling effect.  Note that a more sparsely sampled model grid in $T_{\mathrm{irr}}$ and $g$ was used in the compact and nucleation-limited simulations (see Table~\ref{table:gridparams}). Additional temperature profiles showing profiles at all latitudes and longitudes for selected models are provided in Appendix \ref{sec:appendix_tp}.}  
    \label{fig:t_p_cond}
\end{figure*}

As a result of this range in cloud distributions, the degree to which the potentially observable fluxes from these cloudy atmospheres will deviate from the corresponding clear atmospheres is strongly temperature dependent. If clouds are vertically extended and at least mildly absorbing, colder planets will have relatively enhanced dayside fluxes at both visible and thermal wavelengths due to the visible scattering and absorption. 
In contrast, the cloudy nightsides, receiving no stellar heating, will have suppressed thermal emission as clouds trap heat at depth, as can be seen in the top row of Figure \ref{fig:cloudgrid_night} and temperature profiles in the second row of Figure \ref{fig:t_p_cond}.  As the irradiation temperature increases, the clouds dissipate and the effects are reduced, but to different extents depending on the precise temperatures.  These general mechanisms will dictate the trends seen in the phase curves as discussed in Section \ref{phasesection}.

\begin{figure*}
    \centering
    \includegraphics[scale=0.78]{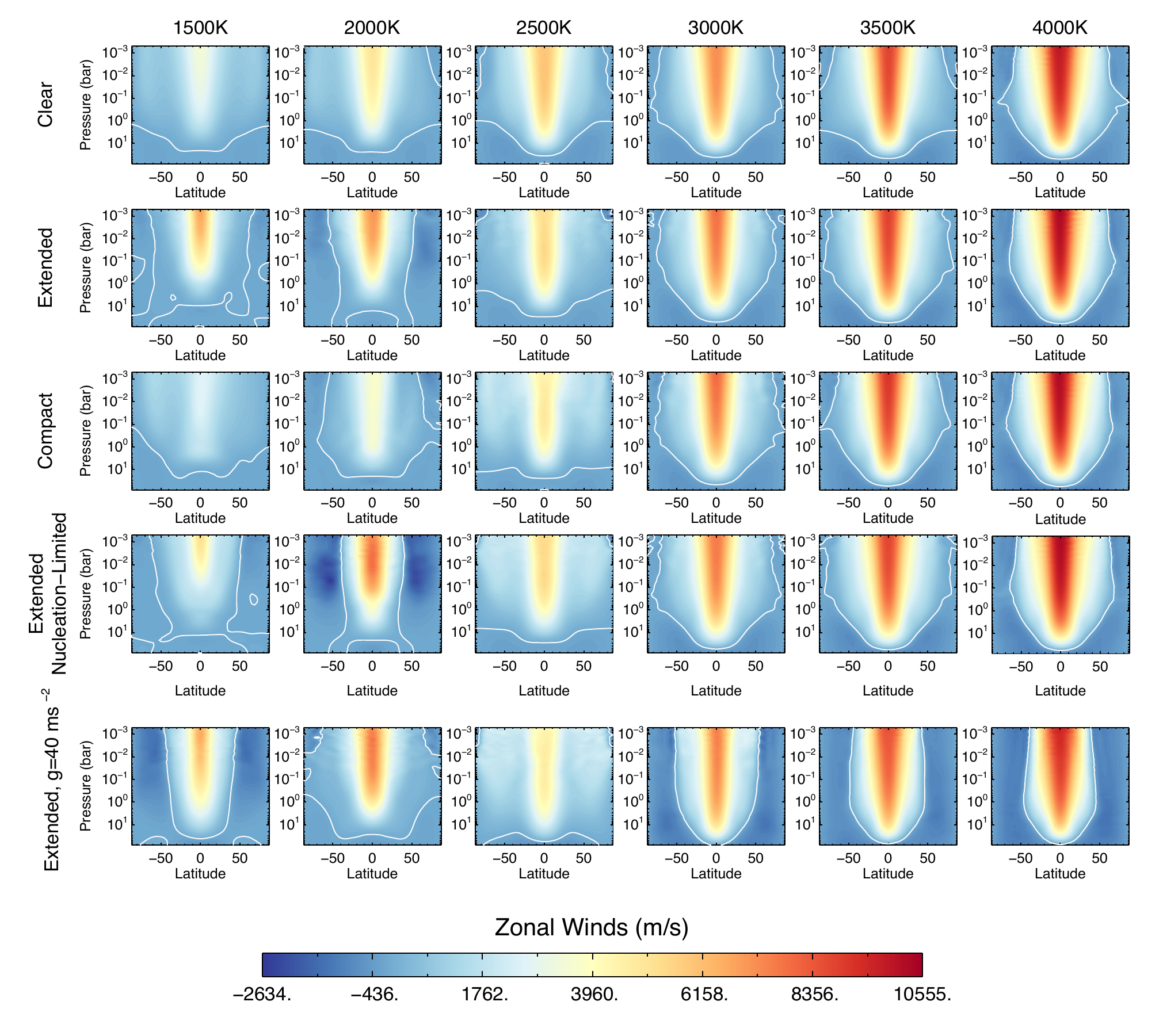}
    \caption{Zonally averaged winds as function of pressure and latitude at six irradiation temperatures for a selected range of cloud modeling assumptions. Positive values show eastward (prograde) winds, while the white contour marks 0 m/s. The longitudinally-averaged eastward equatorial jet and westward high-latitude winds increase with increasing stellar forcing on the dayside. When clouds are present and vertically extended, the equatorial jet becomes stronger in the cooler models as visible scattering and absorption across the dayside increases the day-night and equator-to-pole temperature gradients that ultimately transfer momentum to the equatorial jet. As the irradiation temperatures increases, clouds then preferentially form along the limbs and nightside, thinning along the equator, reducing temperature and wind gradients. As temperature increase further, the effect of clouds diminishes. By comparison, the vertically compact clouds have relatively less effect on the dynamics than the extended clouds, while the nucleation-limited clouds can produce stronger jets when clouds distributions preferentially reduce absorption at the poles (e.g. at T$_{\mathrm{irr}}$ = 2,000 K). Results are shown for 10$\%$ cloud mass and $g=10$ m s$^{-2}$, except for the bottom row which shows that a relatively more narrow and vertically extended jet results when $g=40$ m s$^{-2}$. Trends in corresponding cases for 100$\%$ cloud mass and $g=40$ m s$^{-2}$ are qualitatively similar and are not shown. }
    \label{fig:zonal_all}
\end{figure*}

Combined with their thermal and radiative effects, the clouds in our models also influence atmospheric dynamics, as seen in zonally-averaged winds shown in Figure~\ref{fig:zonal_all}.  In general, wind speeds increase with irradiation temperature, as is typical for cloud-free hot Jupiter GCMs. Strong heating on the dayside leads to strong day-night temperature contrasts that create waves and transfer momentum to the prograde equatorial jet \citep{ShowmanPolvani2011}.  However, when clouds are present, the heating pattern is disrupted and the resulting dynamics are altered. In the cooler models, extended clouds scatter and absorb visible radiation across the dayside, increasing the day-night and equator-to-pole temperature gradients and significantly altering the vertical profiles and atmospheric stability. Consequently, at the coldest irradiation temperatures, the equatorial jet in the extended cloudy simulations is stronger than in the corresponding clear simulations and winds even change direction to become weakly westward at high latitudes.  As the irradiation temperature increases, the heating distribution changes as clouds preferentially form along the limbs and nightside, thinning along the equator, weakening temperature gradients and leading to a slightly weaker jet and more eastward winds at high latitudes.  As temperatures increase further, skies become clearer and the effect of clouds on the heating rates and dynamics diminishes.

%\clearpage

%\null
\subsubsection{Role of Different Cloud Species}\label{sec:species}
The cloud behavior in our baseline model grid is brought on by a combination of the 13 individual cloud species that we consider. In our extended cloud models, many of these species tend to form together, with a mixture of optical properties determined by the relative optical thickness of each species. The optical depth contributions from each of these cloud species are shown in Figure~\ref{fig:cloud_maps} for the case of a planet with $T_{\mathrm{irr}} =$ 2,250 K and $g = 10$~m~s$^2$, and in Appendix~\ref{sec:appendix_cloud_maps} for various planets spanning our full model grid. Some of the cloud species play only a minor role due to their low abundances (e.g.\ VO) or low condensation temperatures (e.g.\ KCl, ZnS, and Na$_2$S) that cause them to only form over limited regions of the very coldest planets in our model grid.  In the colder cases, Cr$_2$O$_3$ serves as a significant source of radiative heating as it forms low-albedo clouds with modest opacity.  Likewise, Al$_2$O$_3$, with a greater abundance and the highest condensation temperature of all our expected cloud species, is a significant absorber at visible wavelengths in even the warmer extended models. However, in most cases, the cloud opacity and aerosol properties of all these species are dominated by SiO$_2$, Fe, and Mg$_2$SiO$_4$, owing to their considerably greater abundances (see Table~\ref{table:clouds}). 

The abundant silicates --- SiO$_2$ and Mg$_2$SiO$_4$ --- each produce thick clouds of conservatively scattering particles at similar temperatures that, when combined, dominate the clouds by mass. (Note that in our models we elect to form forsterite (Mg$_2$SiO$_4$) rather than enstatite (MgSiO$_3$) because the former condenses at slightly higher temperatures, but this choice should have little bearing on our results, as both species have very similar optical properties). In contrast, iron cloud particles have relatively low single scattering albedos and extremely high extinction efficiencies in both the visible and IR (for details, see Figure~\ref{fig:scatparams} in Appendix \ref{sec:appendix}). As a result, Fe is the dominant absorbing cloud in these extended cloud simulations and thus the greatest contributor to the visible heating rates and thermal opacity while the silicates are the dominant conservative scatterers.  Since iron clouds condense at temperatures only slightly greater than the silicate clouds, these three most massive clouds form at similar temperature and pressures, resulting in a cloud with blended properties. Typically, we find the large mass of iron is enough to reduce the overall albedo of the aggregate cloud in our modelling, thus making the overall cloud more absorbing. Therefore, if present in the visible atmosphere, iron clouds likely play a major role in the radiative heating and planetary albedo.  But given the relatively lower nucleation rates of iron as a condensate, doubt has been cast on its theoretical likelihood \citep{gao2020aerosol}.  Furthermore, all three of these abundant species are relatively warmer clouds that would condense quite deeply in the atmosphere of a colder planet, and therefore would only be visible at the top of the atmosphere if mixing produces vertically extended clouds. In Section~\ref{sec:3.2} we discuss what happens when each of these physical assumptions (e.g.\ pure chemical equilibrium and vertically extended clouds) is called into question.

\begin{figure*}[t!]
    \centering
    \includegraphics[scale=0.6]{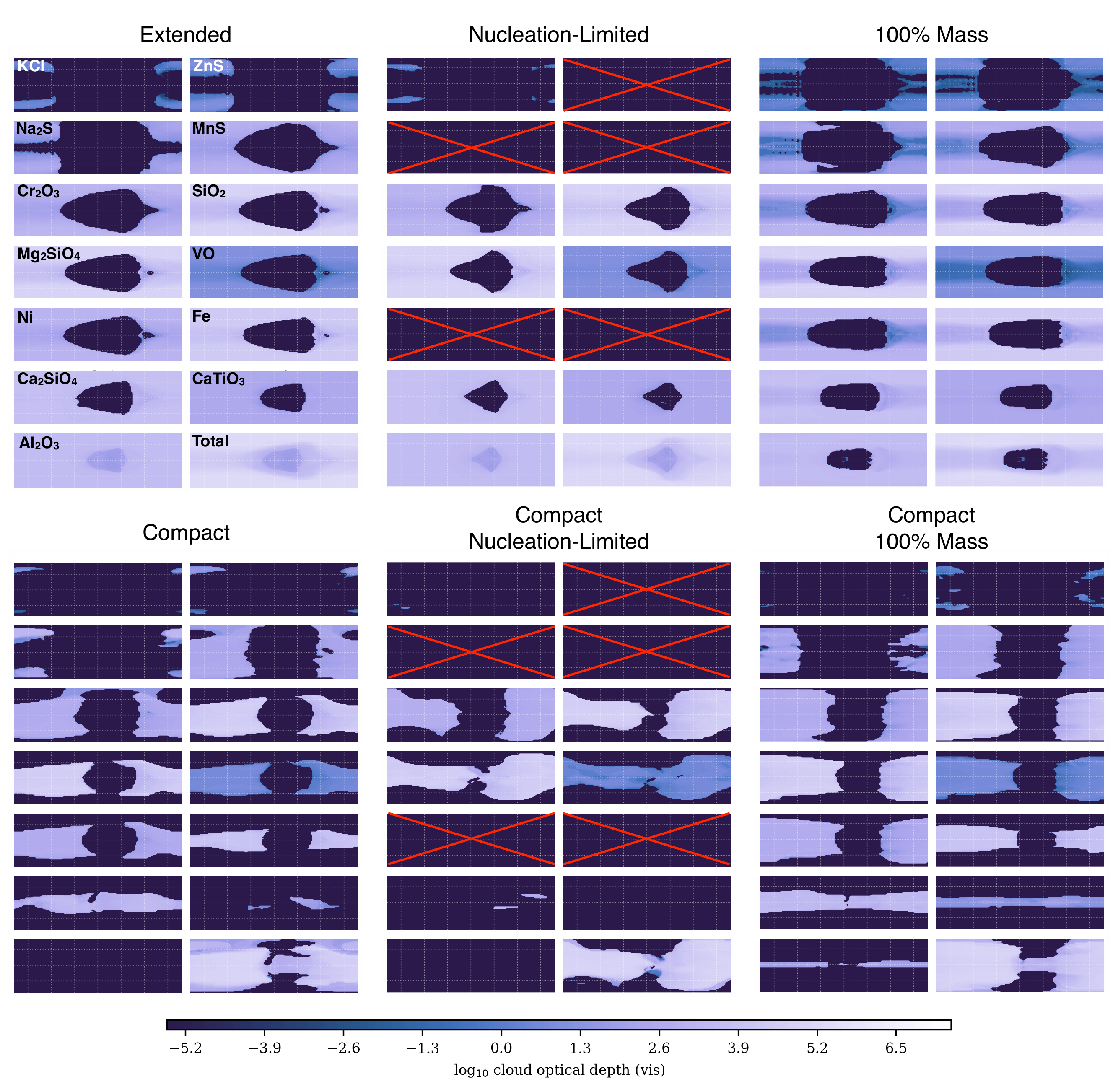}
    \caption{Species-by-species cloud maps showing the visible-wavelength optical thickness (above the cloud-free visible photosphere) for each individual cloud species in the case of a planet with $T_{\mathrm{irr}} =$ 2,250 K and $g = 10$ m s$^{-2}$.  Outcomes for all six cloud implementations are shown, as indicated by labels at the top of each panel.  Labels indicating which cloud species corresponds with which map are shown only in the upper left-hand panel, but are the same for the other panels.  In all cases, the lower right-hand map shows the net optical depth from summing all 13 cloud species.  In the nucleation-limited cases, clouds of Fe, Ni, MnS, Na$_2$S, and ZnS are not included in the GCM, as indicated by red ``x'' signs.  Clear differences can be seen between the cloud maps for the six different cloud models, with the most prominent changes occurring between models with extended and compact clouds.  More subtle differences among the three extended cloud (or compact cloud) cases result from the radiative feedback brought on when different clouds form. We highlight the case of $T_{\mathrm{irr}} =$ 2,250 K in this figure because of the particularly strong differences seen between the various cloud implementations.  Species-by-species cloud maps for models over a broader range of $T_{\mathrm{irr}}$ are shown in Appendix~\ref{sec:appendix_cloud_maps}, for completeness.}
    \label{fig:cloud_maps}
\end{figure*}

\subsubsection{Dependency on Surface Gravity}
The effect of surface gravity on our results is secondary compared to other factors investigated.  Generally speaking, higher surface gravity planets have moderately higher-pressure photospheres, allowing stellar radiation to penetrate and warm deeper layers.  As a consequence, cloud layers also form deeper in the atmosphere because intersections of the vertical temperature-pressure ($T$-$P$) profiles with the near-isothermal condensation curves tend to occur where temperatures increase more rapidly with depth (see Figure~\ref{fig:t_p_cond}). Since the potential cloud and gas masses both scale equally with gravity, the amount of aerosol seen above the photospheres remains similar in each case, although the stellar heating will be occurring at greater pressures.
However, since the cloud particle sizes are prescribed in our model based on the layer rather than the pressure, irrespective of gravity, the precise particle scattering parameters encountered between the photospheres and the top of the atmosphere can differ. The effect is apparently minimal but can explain some differences in heating and scattering at transitional temperatures. It is worth stressing, however, that in real atmospheres, we would expect that the particle size and vertical distribution of aerosols would be dependent on the surface gravity as larger particles would settle more quickly with reduced aerosol scale heights. As a result, in the highest gravity cases, we may expect to see clouds with effects more akin to those found in the compact model simulations rather than the extended cloud simulations. 

The choice of surface gravity has a more significant effect on the dynamics (Figure~\ref{fig:zonal_all}). %The overall larger atmospheric momentum combined with the greater photospheric depths lead to differences in zonal winds.  
The prograde (eastward) equatorial jets in the higher gravity cases are marginally weaker, narrower, and extend deeper, consistent with stellar heating reaching greater pressures. At high latitudes, winds remain relatively variable but become more westward in zonal averages in the \mbox{$g=40$~m~s$^{-2}$} cases.  The reduced width of equatorial jets and the increased westward tendencies of zonally averaged winds at higher latitudes are likely consequences of the the lower atmospheric scale height (inversely proportional to $g$), which in turn modifies the shape of the flow by reducing the Rossby deformation radius \citep{ShowmanPolvani2011}. The overall changes are still minor though, with little effect on the temperature fields and resulting cloud distributions.

\subsection{Dependencies on Cloud Model Implementation} \label{sec:3.2}

As seen in Figures~\ref{fig:cloudgrid}-\ref{fig:zonal_all}, the expected effects of clouds in hot Jupiter atmospheres can vary significantly depending on basic assumptions regarding their vertical distribution and composition. 

\subsubsection{Compact Clouds}\label{sec:compact}
Whether clouds contribute to the scattering of incident stellar light or alter the intensity and transmission of thermal emission strongly depends on the cloud abundance and visual properties above the optical and thermal photospheres.  If the atmosphere is assumed to be vigorously mixed such that cloud particles are lofted far above their initial condensation heights and present anywhere thermochemical equilibrium permits (as in our nominal models) then clouds can contribute towards the planetary albedos and radiative heating rates regardless of the pressure at which they first condense. But if clouds are vertically limited due to weaker vertical mixing and the consequent reduction in available vapor abundances with height (as expected in the Solar System's gas giants), then the height at which they first condense will determine the effect on the observable atmosphere.  Our compact cloud models evaluate this latter assumption.

Given a reduced vertical extent, our compact clouds models have less cumulative opacity than the corresponding extended cloud models; consequently, the radiative impact of the clouds is relatively reduced compared to the extended cloud models.  The resulting atmospheric temperatures, winds, and radiative fluxes are, in most cases, more similar to the clear-atmosphere models, as  \citet{Roman&Rauscher2019} had found in their modeling of Kepler-7b. But, interestingly, with our grid spanning a range of irradiation temperatures, we find that the effects of compact clouds can vary greatly depending on atmospheric temperatures.  Since the compact clouds in our models only extend upward a single scale height's distance above the cloud base, only those clouds forming near or above the visible photosphere will be seen and contribute to the absorption and scattering of instellation.  In colder cases, nearly all species can become trapped below the visible atmosphere, resulting in largely cloud-free conditions in the visible atmosphere, as seen in the $T_{\mathrm{irr}}=1,500$ K case (see Figure \ref{fig:cloudgrid}).  

As temperature increases across our model grid, a sequence of different compact clouds sweep into and out of view at the visible photospheres of our models, forming broken cloud coverage and separating species that would otherwise be mixed together in our extended models. For this reason, our compact cloud models display much more diversity in albedo and thermal properties as a function of $T_{\mathrm{irr}}$ than our extended cloud models.  For example, in the $T_{\mathrm{irr}}=2,500$~K compact cloud case, the slightly deeper and warmer iron cloud becomes concealed by the overlying silicate cloud nearer the colder limb, producing abrupt changes in cloud albedo between the clear center and cloudy limb (see Figure \ref{fig:cloudgrid}).  The pattern is remarkably different, however, at $T_{\mathrm{irr}}=2,250$ K, where the atmosphere is cold enough at low latitudes to form iron clouds on the eastern limb, but too cold to form iron clouds in the visible atmosphere at higher latitudes, instead revealing brighter and colder silicate clouds (see Figure \ref{fig:cloudgrid}). Such examples clearly illustrate how significant and varied observations can be if clouds form distinctly separate layers, as naturally may be expected if vertical mixing is weak and clouds become vapor-limited with height.

\subsubsection{Extended Nucleation-Limited Clouds}\label{sec:extnuc}

As discussed in \ref{sec:species}, we find that, given their great mass and ability to strongly absorb, clouds composed of iron dramatically shape the radiative heating rates and aggregate cloud albedos in our baseline models. In the nucleation-limited models, iron clouds are forcibly absent, along with ZnS, Na$_{2}$, MnS, and Ni, owing to the assumed high nucleation barrier to forming these species.  The lack of massive iron clouds allow conservatively scattering silicate clouds to overwhelmingly dominate the optical properties, producing thick, bright clouds that reflect much of the incident stellar radiation while still suppressing thermal emission over the entire planet, compared to clear-atmosphere models. Without the absorptive iron clouds, dayside aerosol radiative heating is greatly reduced, leading to much weaker thermal inversions and smoother temperature profiles in the limited-nucleation models (Figures~\ref{fig:t_p_cond} and~\ref{fig:tinversion}) compared to our baseline cloud model, although some aerosol heating remains due to the presence of more modestly absorbing clouds (primarily Al$_2$O$_3$).  As a result of the radiative feedback included in our simulations, the reduced absorption also results in thicker, more uniform cloud coverage, which, being considerably more reflective across the dayside, increases the overall planetary albedo and consequently lowers global temperatures and thermal emission compared to the extended cloud case (see Figure \ref{fig:cloudgrid}). 

\subsubsection{Compact Nucleation-Limited Clouds}\label{sec:comnuc}
Limited in both vertical extent and composition, these clouds have effects consistent with qualities of each of the aforementioned models. The lower cumulative opacity generally reduces the effect on the temperature structures relative to the extended cases. At colder $T_{\mathrm{irr}}$, confined to nearer the condensation level, clouds have relatively little to no effect on the albedo and thermal fluxes compared to the corresponding extended cloud models. In the $T_{\mathrm{irr}}=$ 1,500 K case, lacking strongly absorbing clouds in the visible atmosphere, the atmosphere cools enough to appear mostly cloud free on the dayside. However, the constraint on vertical thickness as discussed in Section \ref{sec:compact} again results in a rich but slightly different variety in spatial distributions. With iron and nickel clouds absent and Al$_2$O$_3$ forming at warmer depths below the visible atmosphere, the silicates are free to form pure and highly reflective clouds at temperatures found in the observable atmosphere at moderate irradiation temperatures. In the $T_{\mathrm{irr}}=2,500$~K case, this results in a crescent of exceptionally bright clouds along the western limb, more so than is seen in the other cloud models, while at $T_{\mathrm{irr}}=2,250$~K, this yields to unusual scenario in which highly reflected clouds are seen only on the \emph{eastern} limb. This transition ultimately produces an anomalous eastward offset in the reflected-light phase curve for our $T_{\mathrm{irr}}=2,250$~K model as discussed in Section \ref{phasesection}, and again illustrates how subtle differences in atmospheric temperatures can result in a varied array of observed phase curves if clouds are limited in vertical extent. 

\subsubsection{Effect of Cloud Mass}\label{toomanyclouds}
For our standard simulations, we assume that 10$\%$ of the vapor, by mass, condenses to form clouds. Simulations repeated assuming 100$\%$ show a qualitatively similar but more intense response. The factor of 10 increase in mass directly yields greater cloud opacities, radiative heating rates, and resulting feedback in the cloud distribution.  At lower irradiation temperatures, this leads to increases in the absorption and thermal emission on the daysides and further reduction in the thermal emission from the nightside. At intermediate temperatures, the intensity of the radiative heating causes clouds to be even patchier while inversions grow even stronger. From a practical perspective, the extreme heating rates in these upper-limit cases appear to challenge the numerical stability of the models, requiring us to introduce the stabilizing measures as discussed in Section \ref{sec:cloudmodel}. As can be seen in dayside differential fluxes and thermal inversions (Figures \ref{fig:cloudgrid} and \ref{fig:tinversion}, bottom rows), slight numerical artifacts in the form of square waves in temperatures fields begin to emerge in the cloudiest cases. These artifacts are related to cloud transitions within the resolution of the spectral model's Gaussian grid and evidently do not have any significant effect on the cloud distribution or dynamics, which appear in line with the trends exhibited in lower cloud mass cases.  Altogether, this shows that though increasing the cloud mass will amplify the expected effects of clouds, the general trends are robust whether 10$\%$ or 100$\%$ of the vapor is assumed to condense, and any differences are minor compared to the assumptions regarding the vertical extent and composition of clouds.

\subsubsection{The Effects of Different Cloud Models on Dynamics}
With respect to dynamics, among the different cloud models, the most pronounced differences exist between the compact and extended cloud models at colder temperatures.  This is simply a result of the compact cloud models at lower $T_{\mathrm{irr}}$ having nearly clear visible photospheres, such that there little to no differences in heating relative to the a clear atmosphere.  Accordingly, these compact cases have winds that more closely resemble the clear cases rather than the corresponding extended cases, whether nucleation-limited or not (see Figure \ref{fig:zonal_all}). As the clouds condense higher in the slightly warmer cases (e.g.\ $T_{\mathrm{irr}}=$ 2,500 K), the wind field becomes more like the extended cases.  Differences between the baseline extended and nucleation limited cases are more subtle, with a modest increase in the magnitude of both eastward and westward winds in the cool, completely clouded cases (e.g.\ $T_{\mathrm{irr}}=$ 2,000 K), and only a slight increase in mid- and high-latitude eastward winds in the warmer, partly cloudy cases (e.g.\ $T_{\mathrm{irr}}=$ 2,500 K). Overall, these differences and their consequence on observable quantities is minor compared to the significant and direct effects the clouds have on the observed albedos and phase curves, as discussed in Section~\ref{phasesection}.

\subsection{Implications for Phase Curve Observations}\label{phasesection}

\begin{figure*}[th!] 
\centering 
\includegraphics[clip, trim=0.2in 0.2in 0.2in 0.2in,width=\textwidth]{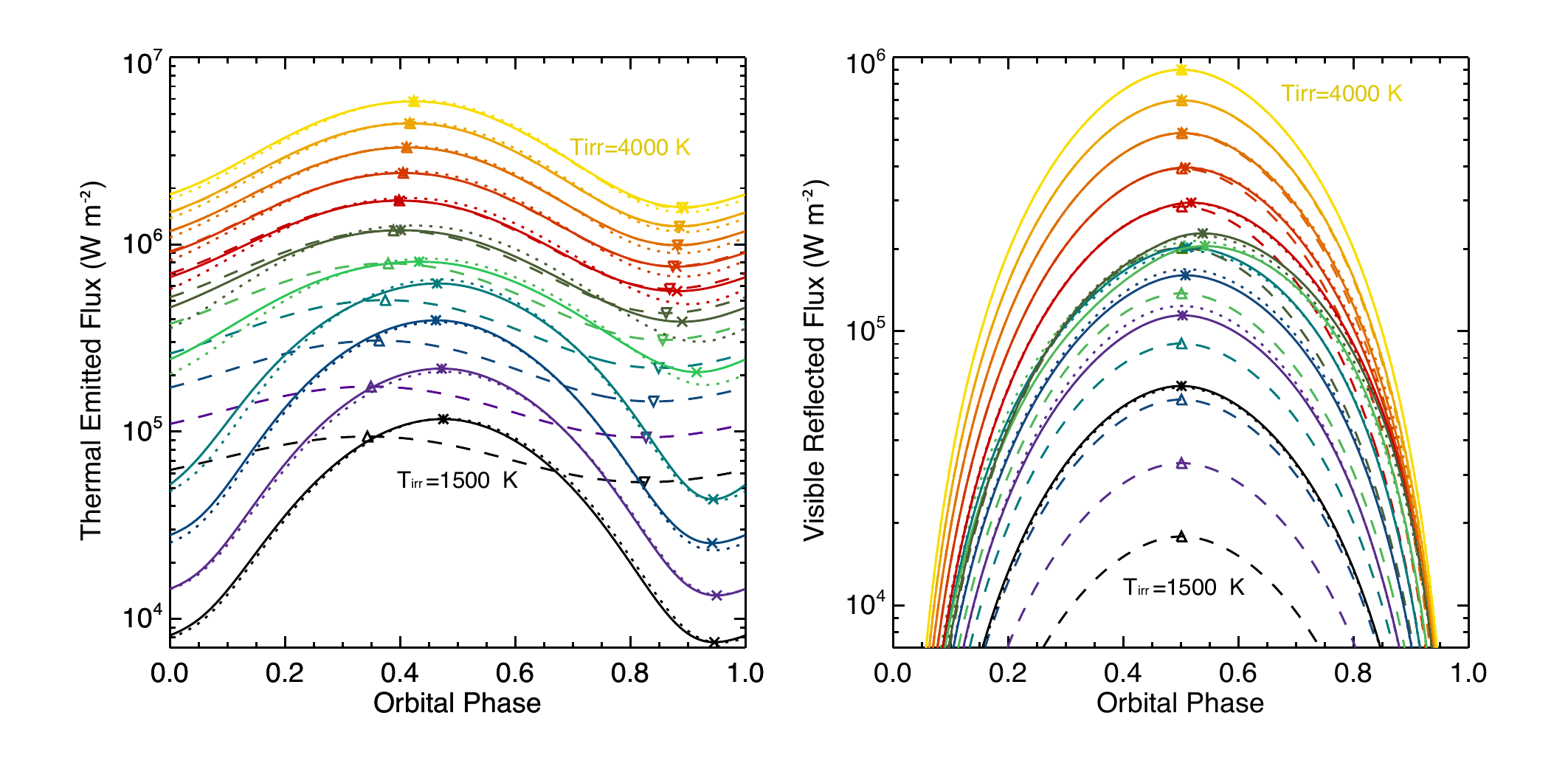}
\caption{Thermal and reflected-light phase curves produced by the different simulated cases. Left panel: The disk integrated thermal emission as a function of orbital phase for clear cases (dashed lines) and extended cloudy cases (solid lines) for irradiation temperatures ranging from 1,500 K (lower black curve) to 4,000 K (upper yellow curve),  assuming $g=10$ m s$^{-2}$. The dotted lines represent the corresponding curves for cloudy cases with $g=40$ m s$^{-2}$. The maximum and minimum fluxes are marked for the clear cases (triangles pointing upward and downward, respectively) and cloudy cases (``$\ast$'' and ``$\times$''-signs, respectively) for $g=10$ m~s$^{-2}$ (also see Figure \ref{fig:Offsets}). For $T_{\mathrm{irr}} \lesssim$ 2,750 K, extended clouds significantly suppress emission on the nightside and enhance dayside emission. Right panel: Reflected light phase curves for the same set of models. As in the left panel, cloud models assuming $g=40$ m\ s$^{-2}$ are also shown (dotted lines) but are difficult to make out because they tend to be nearly coincident with the $g=10$ m~s$^{-2}$ cases. The reflected visible fluxes predictably increase with increasing $T_{\mathrm{irr}}$, but clouds significantly raise the reflectances at cases with $T_{\mathrm{irr}} \lesssim 2,750$ K. Curves for $T_{\mathrm{irr}}=$ 2,500 and 2,750 K (solid light and dark green curves, respectively) have notable asymmetry due to clouds preferentially forming along the western limb at transitional temperatures.}
\label{fig:phasecurves}
\end{figure*}

The thermal and reflected-light phase curves for our clear and standard extended cloud models are shown in Figure~\ref{fig:phasecurves}, illustrating how the trends in temperature and clouds affect these disk-integrated values. Phase curves are calculated by integrating the outgoing flux across the planet's disk at 1,000 different viewing geometries in each channel of our double-gray model. In this framework, the visible (i.e.\ scattered light) and IR (i.e.\ thermal emission) channels are treated separately, so any potential contribution from the thermal emission at visible wavelengths is ignored at this time, as discussed in Section~\ref{sec:discussion}.  At high temperatures where few if any clouds form, the phase curves produced by the cloudy and clear models are nearly identical with peaks in the thermal emission offset from center (i.e. shifted to the east).  As $T_{irr}$ decreases, the thermal offsets in the clear models become greater as the amplitudes of the curves decrease, as is typical of hot Jupiter GCM results \citep[e.g.][]{KomacekShowman2016}.  But in the cloudy models, as temperature decreases, clouds become increasingly prominent, and differences between the two cases become more stark. Relative to the corresponding clear cases, the peaks in thermal emission from the cloudy cases marginally increase while the minima dramatically decrease. The phase of the peaks and minima also shift toward center, such that the cloudy atmospheres produce phase curves with larger amplitudes and smaller peak offsets with decreasing irradiance --- opposite of the trend seen in the clear cases.  

In reflected light at visible wavelengths, cloudy planets are brighter --- by up to a factor of two or more.  Westward peak offsets are predicted for some of the models in the intermediate $T_{\mathrm{irr}}$ range, where we find that cloud distributions are strongly skewed toward the western limb.  However, no significant phase offsets are expected at irradiation temperatures less than 2,250 K, where clouds tend to form more uniform distributions across the disk in our baseline extended models. 

\begin{figure*}[p] 
\centering 
\includegraphics[clip, trim=0.5in 0.in .5in 0.0in,scale=0.75]{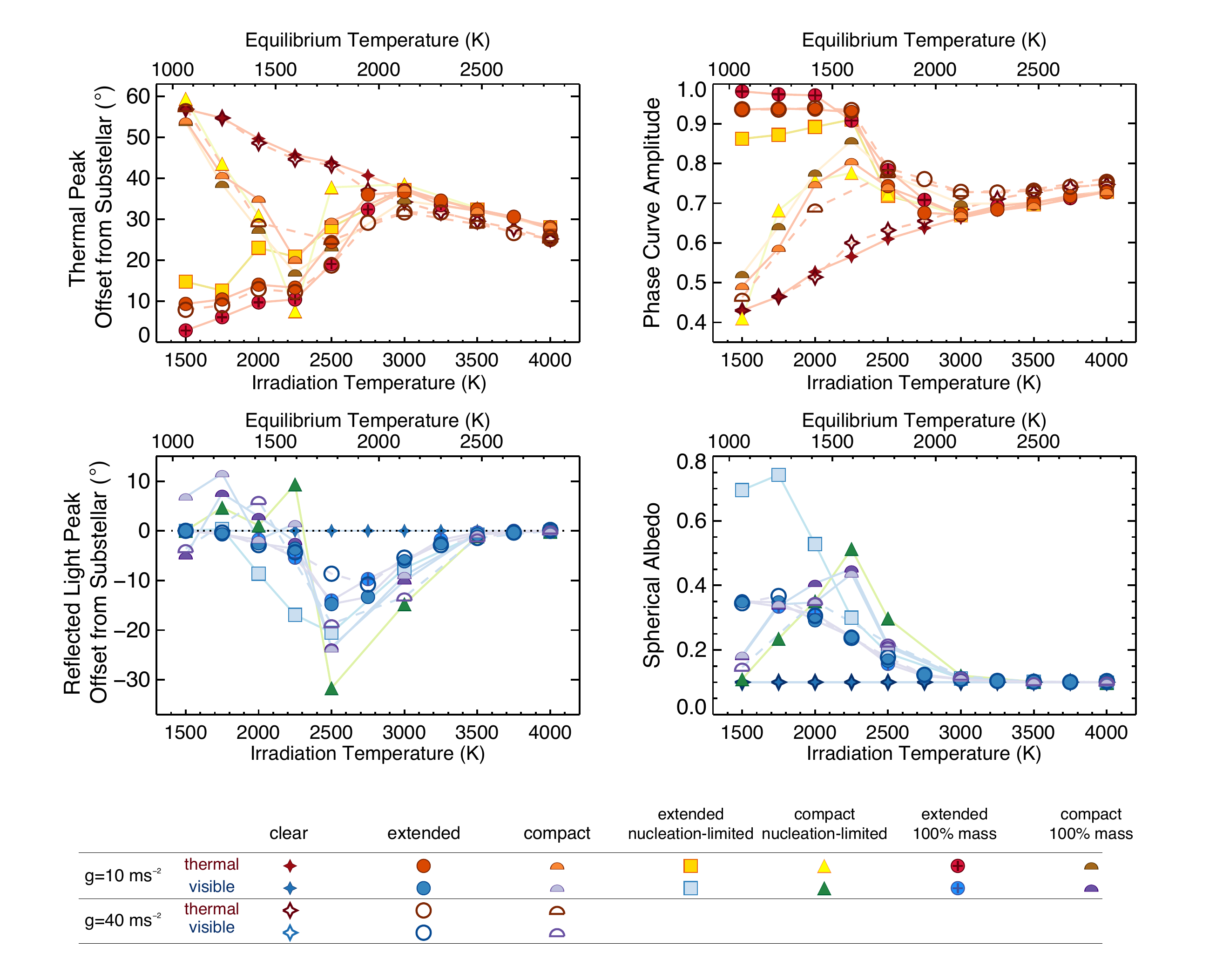}
\caption{Disk-integrated quantities for all simulations as a function of irradiation temperature. Cases are marked by the different symbols listed in the key.  (Left) The longitudinal offsets of the disk-integrated maximum thermal emission (upper left) and reflected light (lower left) relative to the substellar longitude are plotted as a function of irradiation temperature.   (Corresponding equilibrium temperatures assuming zero albedo and fully efficient heat transport are also provided on the upper x-axis). For clear models, the thermal emission peaks to the east of the substellar longitude (positive values), decreasing roughly linearly with increasing irradiation temperatures, while the reflected light peak remains centered at the substellar point. When clouds are present, the eastward thermal shifts are suppressed, roughly converging with the clear cases at $T_{irr} \approx$ 3,000 K. Conversely, in the reflected light (optical) phase curves, clouds forming only in cooler western regions produce westward shifts at irradiation temperatures between 2,250 and 3,250 K. (Upper Right) The corresponding amplitude of thermal phase curves, taken as the difference between the maximum and minimum disk-integrated thermal fluxes divided by the maximum.  Amplitudes are increased in the cloudier simulations as the peak offsets and nightside fluxes are significantly reduced and photospheres move upward in the atmosphere where radiative timescales are shorter. (Lower Right) The spherical albedos, calculated as the fraction of incident stellar radiation diffusely reflected over the disk,  plotted for each case. Clouds raise the spherical albedo well beyond the Rayleigh scattering baseline of the clear models. The nucleation limited cases have considerably higher albedos primarily due to the absence of Fe clouds.}
\label{fig:Offsets}
\end{figure*}

For the range of alternative cloud models investigated, we find a diversity of phase curve behaviors. The offset and amplitude of the phase curves for all simulations are summarized in Figure~\ref{fig:Offsets}. From these plots, the relative significance of different modeling assumptions is evident. Firstly, clouds begin to produce diverging behavior from clear atmospheres below $T_{\mathrm{irr}} =$ 3,000~K, with extended clouds tending to produce the most extreme differences in thermal offsets and amplitudes.  In all cases, clouds consistently lead to larger phase curve amplitudes and smaller peak offsets than in the clear-atmosphere case.  However, the magnitude of these effects varies with the cloud model employed.  Of the extended cloud atmospheres, the 100\% condensed case produces the most dramatic deviations from the clear-atmosphere simulations, followed by the baseline and nucleation-limited cases.  The compact clouds tend to mimic the phase curve behavior of the extended clouds at intermediate $T_{\mathrm{irr}}$, before converging back toward the behavior of the clearer models at the coldest irradiation temperatures.  The thermal phase curve properties are only marginally sensitive to the choice of surface gravity --- and more so for the compact models than for the extended ones.

These thermal phase curve behaviors can be attributed to effects that have been discussed above. On the dayside, thick extended clouds absorb heat and increase opacity, thereby raising the height and temperature of the photosphere. This effectively reduces the photosphere's radiative time constant \citep{iro2005timeconstants,seager2005dayside,CowanAlgol2011}. On the nightside, clouds suppress the thermal flux from below, which, when accompanied by a higher-altitude and colder photosphere, dramatically reduces the thermal emission. This results in very large phase curve amplitudes and small peak offsets in the colder cloud-dominated models, bucking the expected trend for clear atmospheres that tend to have the reverse behavior as a function of $T_{\mathrm{irr}}$.  The nucleation-limited models show similar behavior to the extended cloud case, but it is less dramatic because of the much weaker dayside heating and thermal inversions.  The compact cloud models vary between behavior more akin to that of the clear or extended cloud models, depending on the exact level of irradiation received by the planet.  This occurs as clouds move into and out of view at the visible photosphere, similar to behavior identified by \cite{parmentier+2016}.  For example, at $T_{\mathrm{irr}} =$ 1,500 K, metal and silicate clouds are buried deep in the atmosphere well below the visible photosphere, and the planet remains mostly too hot for KCl, ZnS, and Na$_2$S clouds to form, resulting in comparable phase curve behavior to a near-clear atmosphere.

At visible wavelengths, planets with $T_{\mathrm{irr}} <$ 3,000~K are  brighter due to enhanced scattering in our cloudy models, as seen in the albedo maps of Figure \ref{fig:cloudgrid} as well as the integrated spherical albedos in Figure \ref{fig:Offsets} (bottom right panel). The largest albedos are produced by our extended nucleation-limited cloud models, which intercept and scatter light most conservatively. By the same token, the hottest and clear-atmosphere cases produce the smallest albedos, with only molecular Rayleigh scattering countering the gaseous absorption. In between these extremes, we see a range of values depending on the combinations of vertical extent, composition, and spatial coverage.  For example, of all the cloudy cases, the coldest compact-cloud cases produce notably low albedos because the bulk of the cloud mass condenses below the visible atmosphere. 
Visible phase curve peak offsets will sit right at 0$^{\circ}$, unless the surface brightness profile is inhomogeneous, as occurs when the atmosphere is only partly cloudy.  We find westward phase curve offsets are obtained for all cloudy simulations with \mbox{2,250~$< T_{\mathrm{irr}} <$ 3,000~K}.  At these intermediate temperatures, reflective clouds form only on the nightside and western limb, while the atmosphere remains clear on the hotter eastern limb and dayside. At higher temperatures the planets are entirely cloud-free, and at lower temperatures they are either fully enshrouded in clouds or cloudy at greater depths, leading to reflected-light phase curves that are more or less symmetric about the sub-stellar point.

In contrast to the IR phase curve behavior, reflected-light phase curves produced by clouds that are either compact and/or nucleation-limited display more extreme behavior than the extended cloud models, due to the higher albedos that result in these cases.  For both the compact and nucleation-limited models, the dominance of reflective silicate clouds leads to larger westward peak offsets predicted at intermediate temperatures.  Interestingly, several of the compact models with $T_{\mathrm{irr}} \leq$  2,250 K produce small \emph{eastward} offsets at optical wavelengths.  The reason for this can be seen in the albedo maps in Figure~\ref{fig:cloudgrid}, which show multiple cloud species preferentially forming on the eastern dayside in these cases. This is most strongly evident in the compact nucleation-limited case at $T_{\mathrm{irr}} = $ 2,250 K, where the massive silicates (SiO$_2$ and Mg$_2$SiO$_4$) form thick reflective clouds along the eastern limb. Though these precise results are sensitive to the particular parameters of our models, this illustrates how a wider diversity of phase curve offsets and amplitudes can be produced if clouds have relatively limited vertical-length scales in atmospheres with strong east-west gradients in temperature. 

\section{Discussion and Conclusions} \label{sec:discussion}

We have produced a novel set of cloudy 3-D hot Jupiter models to study the impact of clouds as a function of irradiation and surface gravity on thermal structures and phase curve observables.  With approximately half of the \textit{Spitzer} phase curve observations currently published in the peer-reviewed literature, and the rest currently being analyzed, as well as a multitude of optical phase curves expected from \textit{TESS}, statistical trends within the population are being established, and there is accumulating evidence that clouds play an important role at lower planetary temperatures.  In our current work we have investigated how condensate clouds will alter the classic cloud-free hot Jupiter GCM predictions.   Our major findings are as follows:

\begin{enumerate}
    \item Condensate clouds are predicted to impact IR phase curve observables for $T_{\mathrm{irr}} \lesssim$ 3,000 K.  At these temperatures, clouds increase thermal phase curve amplitudes by significantly decreasing the nightside emission and, in some cases, increasing the dayside emission, while simultaneously decreasing peak offsets. This corresponds with a decrease in the photospheric pressures.  The magnitude of these effects depends on the details of the cloud model (i.e.\ how vertically extended the cloud is and what cloud species form).

    \item Condensate clouds are predicted to produce westward peak offsets in optical phase curves over a limited range of irradiation temperatures (2,000~K $\lesssim T_{\mathrm{irr}} \lesssim$ 3,500 K), where the cloud coverage has a strong longitudinal dependence.  Within this temperature range clouds are present primarily on the planet's nightside and western limb.  At higher temperatures the observable atmosphere is cloud free, and at lower temperatures the atmosphere is either fully clouded over or relatively clear at visible heights, depending on the vertical extent of the clouds above their base condensation pressures. In some cases, vertically compact clouds can produce eastward peak offsets given predicted longitudinal and vertical temperature gradients.  We note that thermal intrusion into optical wavelengths for the hottest planets may reduce or even reverse the westward phase curve offsets at the higher end of this temperature range --- an effect that is not modeled within the framework of our double-gray GCM.

    \item Radiative feedback from clouds, especially those that are more vertically extended, has a significant impact on the thermal structures of hot Jupiter atmospheres.  Clouds generally suppress nightside temperatures at the thermal photospheric height.  Dayside temperatures can either be enhanced or suppressed relative to clear-atmosphere conditions depending on the effective cloud albedos.  Highly reflective clouds tend to reduce dayside temperatures, whereas more absorptive clouds can produce significant thermal inversions resulting in increased dayside emission.  Of the dominant cloud species produced in our models, silicate clouds have the former behavior and iron the latter. When iron is absent, emission is generally reduced globally at colder irradiation temperatures, but Al$_2$O$_3$ clouds remain as a significant source of radiative heating at warmer irradiation temperatures. 

    \item Surface gravity, while a key planetary parameter, does not have a strong direct effect on the phase curve observables in our models and therefore is unlikely to be the main driver behind the observed scatter in phase curve properties for planets at similar levels of irradiation. However, if gravity significantly affects the vertical structure and/or composition of clouds --- as can only be investigated with detailed microphysical models --- then gravity might indirectly lead to diverse phase curve properties over limited ranges of $T_{\mathrm{irr}}$.  Avenues of future research to understand the diversity of observed phase curve properties should include investigating the additional roles of cloud microphysics, planetary metallicity, chemistry, and/or photochemical haze production.
    
    %Surface gravity, while a key planetary parameter, does not have a dramatic direct effect on phase curve observables in our models and is therefore probably not the main driver behind the observed scatter in phase curve properties for planets at similar levels of irradiation. However, this statement assumes that gravity is decoupled from defining the vertical extent of clouds  (which does produce diverse phase curve properties over limited ranges of $T_{\mathrm{irr}}$) --- an assumption that can only be tested with detailed microphysical models.  Avenues of future research to understand the diversity of observed phase curve properties could include investigating the additional roles of cloud microphysics, planetary metallicity, chemistry, and/or photochemical haze production.
    %probably not the main driver behind the observed scatter in phase curve properties for planets at similar levels of irradiation. While our models suggest clouds may produce a range of results at intermediate temperatures, they cannot explain a potential diversity in observations at colder and warmer irradation temperatures.
    
\end{enumerate}

Our results shed light on the role of radiatively-interacting clouds in hot Jupiter atmospheres as a function of planetary irradiation. However, it is also important to acknowledge the limitations and assumptions of our simplified models in order to understand effects that may have been missed with our current results.  For example, our model grid does not explore the effects of altering the atmospheric composition (i.e.\ the metallicity), rotation period, or internal heat flux.  Of these three parameters, we expect that rotation period is likely to have the least impact as long as the planet remains tidally-synchronized, as the dynamical effect of winds appears secondary to the direct effects of radiative heating in our models.  Any resulting changes in the strength of the equatorial jet, however, could introduce some additional scatter in the predicted offsets in thermal phase curves.

Metallicity and internal heat fluxes are more likely to impact our results due to their direct effect on the thermal structure of the atmosphere, and, in the case of metallicity, by shifting the cloud condensation curves to higher or lower temperatures.  Qualitatively, we expect that intrinsic scatter in the metallicities of hot Jupiter atmospheres (e.g.\ Mansfield et al., in prep) will induce an additional degree of scatter into our phase curve predictions shown in Figure~\ref{fig:Offsets}.  Significantly larger interior heat fluxes, as suggested by \citet{thorngren+2019}, would provide a further source of heating in the lower atmospheres and as a consequence push cloud base locations to higher in the atmosphere.  This would result in reduced cloud optical depths in our extended cloud models and may alter which cloud species are visible at the photospheres for the case of compact clouds, particularly at colder irradiation temperatures.  Making more quantitative statements about the impact of these confounding variables (metallicity, internal heat flux, and rotation rate) requires additional modeling and is beyond the scope of this work.

Our models additionally omit treatments of certain physical and chemical processes that we expect to impact the edge cases of our grid. 
For example, we do not model hazes that may form due to photochemistry and auroral chemistry \citep[e.g.][]{gao2017sulfur} and potentially alter the albedo of the planets.  In practice, haze formation appears to primarily impact hot Jupiter transmission spectra below $T_{\mathrm{irr}} \approx$ 1,350 K \citep{gao2020aerosol}, which lies just below the lower boundary of our modeled parameter range.  We additionally ignore the dissociation and recombination of H$_2$ \citep{bell18, tan19}, magnetic effects \citep{rauscher13,Rogers2014}, and high-temperature gas opacities in the optical that can lead to thermal inversions \citep{parmentier18, lothringer18, gandhi19}.  Since all of these effects are expected to impact our most irradiated models that remain otherwise cloud-free, we point the reader to the various aforementioned references to find cloud-free GCM results and 1-D forward models that include those effects.  

For computational efficiency, our models lack an explicit treatment of cloud microphysics.   We have instead accounted for the poorly constrained unknowns of eddy mixing rates and  coagulation processes by focusing on three broadly defined cloud models --- extended, compact, and nucleation-limited.  We have shown that our results have significant dependencies on the details of the cloud model, so it is important to ask which of these cases is the closest approximation of the physical reality in a typical hot Jupiter atmosphere.

The inclusion of a set of nucleation-limited models to accompany our baseline extended-cloud case was prompted by the results of \citet{gao2020aerosol}, who predict that silicates should dominate over iron clouds due to the relatively low rates of nucleation expected for the iron condensates. Yet, \citet{ehrenreich20} found compelling observational evidence for nightside iron condensation in the transmission spectrum of the ultrahot Jupiter WASP-76b, suggesting that iron may still form clouds, albeit potentially to a lesser degree.  While awaiting future data to further constrain the abundance and opacity of iron condensates, we justify both our nominal equilibrium chemistry and nucleation-limited cloud grids (the former which includes iron clouds and the latter which does not) as limiting cases, illustrating the potential role these clouds may play while recognizing that real hot Jupiter atmospheres likely fall between these two extremes.  Helpfully, our models offer up a new prediction of how dayside iron clouds would significantly reduce the planetary albedo and generate strong thermal inversions over a wide range of irradiation temperatures, providing an avenue for observationally interrogating their presence.

As for the compact vs.\ extended depiction of clouds, we look to the results from cloud microphysics models to inform which one is better aligned with real hot Jupiter atmospheres. The true vertical extent of clouds in a planetary atmosphere is governed by the competing effects of gravitational settling and vertical mixing.  Of these, the vertical mixing rate (parameterized via an eddy diffusion coefficient, $K_{zz}$) is more difficult to estimate.  Various estimates from modeling span orders of magnitude and likely vary with height \citep{[e.g.],Moses2013,Parmentier2013,Agundez2014,lines+2018}. \citet{komacek19} and \citet{powell19} predict values of $K_{zz}$ for hot Jupiter atmospheres ranging from $\sim 10^{5} - 10^{11}$ cm$^2$ s$^{-1}$, with the higher values corresponding to the upper end of the planetary temperature range explored in our own models.   In the cloud microphysics models of \citet{powell+2018, powell19}, clouds tend to be moderately extended --- typically with cloud tops that are $\sim$1 scale height above the cloud base, but in some cases with cloud tops that extend all the way to the top of the modeled portion of the atmosphere, depending somewhat sensitively on the individual cloud species, the local temperature conditions, and the assumed, uncertain eddy mixing rate. Likewise, clouds in the kinetic, microphysics-coupled simulations of \citet{lines+2018} are found to extend from $\mathrm{10^1-10^{-4}}$ bars, truncated by the top of modelling domain in parts, although clouds forming deeper may be more compact due to larger particle sizes and weaker mixing at higher pressures \citep{Lines+2019}. In general though, the clouds produced by the microphysics calculations tend to fall in between the two extremes set by our extended and compact cloud realizations, which leads us to conclude that these should be considered as bounding cases.  

It should also be noted that although the double-gray nature of our GCM calculations provides a relatively computationally efficient way to generate a large grid of cloudy 3-D models, it neglects wavelength-dependent effects that could potentially be important.  \citet{Roman&Rauscher2019} provide a lengthier discussion of the potential impacts of non-gray radiative transfer on cloudy GCM thermal structures.  An additional limitation of our double-gray calculations is that they only allow us to predict broadband thermal and optical phase curve behavior, and they neglect the overlapping nature of these two bands for hotter planets.  This prohibits us from making meaningful \emph{spectral} predictions of phase curve observables --- e.g.\ modeling the different behavior that is seen in \textit{Spitzer} observations obtained at 3.6 and 4.5~$\mu$m or making spectroscopic predictions for \textit{JWST}.  We intend to rectify this shortcoming in a follow-up work in which we will post-process our cloudy GCMs to generate fully wavelength-dependent spectral phase curves that will allow us to directly predict and interpret observations with \textit{Spitzer}, \textit{JWST}, and other facilities (Paper II; Kempton et al., in prep).  

Nonetheless, despite a relatively simplified treatment of cloud physics and chemistry, it is encouraging to note how remarkably similar our results compare to those determined using more detailed modeling in select cases. Our predictions of preferentially cloudy nightsides and western limbs transitioning to clear daysides with rising temperatures follow similar results of \citet{Parmentier2016} and Parmentier et al. 2020 (in press), who post-processed clouds using a model with spectrally-resolved gas opacities and equilibrium chemistry for a grid of planets with $T_{\mathrm{eq}} <$  2,200 K ($T_{\mathrm{irr}} \lesssim$  3,100 K). Likewise, the sensitivity of the emission to the vertical positioning of the cloud, the importance of radiative feedback and its strong effect on cloud distributions, dayside heating rates, and predicted thermal phase curves \citep[in this work and][]{Roman&Rauscher2019} strongly agree with conclusions of \citet{Lines+2019}, who modeled HD 209458b using a more sophisticated cloud parameterization \citep[EDDYSED,][]{ackerman2001precipitating} within a non-hydrostatic GCM with self-consistent chemistry and realistic gas opacities. This suggests that, although relatively simple, our models adequately capture much of the essential physics that shape the thermal structure and cloud distributions in these atmospheres. And given the computational efficiency our approach affords, we are more able to isolate and evaluate the effects of different assumptions over a wider range of conditions with sufficient fidelity. 

Ultimately, the validation of these and all exoplanet models will rely upon comparison to observations. Although generating wavelength-dependent spectral phase curves from our models for comparison is left to future work, we note that our broad-band reflected light curves are likely reasonable approximations of the Kepler passband, assuming the thermal component can be neglected. Therefore we conclude by returning to observations of reflected light phase curves for the hot Jupiter Kepler-7b \citep{latham+2010,demory+2013}. With an irradiation temperature of $\sim$2,300~K, Kepler-7b's phase curve was reported to be offset 41~$\pm$~12$^\circ$ to the west of the substeller longitude. \citet{Roman&Rauscher2017} modeled Kepler-7b with a prescribed silicate cloud distribution \citep{munoz+2015}, but they found that the proposed clouds lacked self-consistency with the temperature field. \citet{Roman&Rauscher2019} then sought to address this inconsistency by finding solutions with temperature-dependent clouds but these simulations fell short in reproducing Kepler-7b's exceptionally high spherical albedo of $0.4-0.5$ \citep{demory+2013,munoz+2015}.  As our current modeling shows, Kepler-7b's irradiation temperature falls within a region of the parameter space where phase shifts and albedos depend strongly on the details of the clouds and their environment, potentially leading to a diverse range of observed values in similar atmospheres.  Now, with additional clouds and various implementations evaluated over a range of temperatures, we find that our models can produce a combination of the visible offset and spherical albedo that agrees with observations of Kepler-7b, but only for one case --- the compact nucleation-limited cloud case at $T_{\mathrm{irr}}$ = 2,500 K (Figure \ref{fig:Offsets}). Although far from conclusive, the better agreement is encouraging and argues for further investigation of the cloud parameters. Similar comparisons between data and models can potentially shed light on the more general nature and diversity of clouds forming in hot Jupiter atmospheres.  Accordingly, interpretation of these and other future observations will continue to benefit from a diversity of cloud modeling.

\acknowledgements
We thank Diana Powell for useful discussions on the microphysical details of cloud formation in hot Jupiter atmospheres.
M.T.R.\ acknowledges support by a European Research Council Consolidator Grant, under the European Union’s Horizons 2020 research and innovation program, grant number 723890.
E.M.-R.K.\ and E.R.\ acknowledge support from the NASA Astrophysics Theory Program (grant NNX17AG25G), the Heising-Simons Foundation, and the Research Corporation through the Cottrell Scholar program.
C.K.H.\ acknowledges support from the National Science Foundation Graduate Research Fellowship Program under Grant No.\ DGE1752814.
J.L.B.\ and K.B.S.\ acknowledge support for this work from NASA through an award issued by JPL/Caltech (Spitzer programs 13038 and 14059).
This research used the ALICE High Performance Computing Facility at the University of Leicester.  The authors wish to acknowledge the significant role that Dr.\ Adam Showman played in laying out the foundational theory of hot Jupiter atmospheric circulation.  We were saddened to hear of his death while we were drafting this manuscript and will miss the insights he would have brought to continued work in this field.

\bibliography{cloudygrid}

\begin{thebibliography}{}
\expandafter\ifx\csname natexlab\endcsname\relax\def\natexlab#1{#1}\fi
\providecommand{\url}[1]{\href{#1}{#1}}
\providecommand{\dodoi}[1]{doi:~\href{http://doi.org/#1}{\nolinkurl{#1}}}
\providecommand{\doeprint}[1]{\href{http://ascl.net/#1}{\nolinkurl{http://ascl.net/#1}}}
\providecommand{\doarXiv}[1]{\href{https://arxiv.org/abs/#1}{\nolinkurl{https://arxiv.org/abs/#1}}}

\bibitem[{Ackerman \& Marley(2001)}]{ackerman2001precipitating}
Ackerman, A.~S., \& Marley, M.~S. 2001, The Astrophysical Journal, 556, 872

\bibitem[{{Ag{\'u}ndez} {et~al.}(2014){Ag{\'u}ndez}, {Parmentier}, {Venot},
  {Hersant}, \& {Selsis}}]{Agundez2014}
{Ag{\'u}ndez}, M., {Parmentier}, V., {Venot}, O., {Hersant}, F., \& {Selsis},
  F. 2014, \aap, 564, A73, \dodoi{10.1051/0004-6361/201322895}

\bibitem[{Al-Kuhaili \& Durrani(2007)}]{al2007optical}
Al-Kuhaili, M., \& Durrani, S. 2007, Optical Materials, 29, 709

\bibitem[{Anders \& Grevesse(1989)}]{anders1989abundances}
Anders, E., \& Grevesse, N. 1989, Geochimica et Cosmochimica acta, 53, 197

\bibitem[{Arduini {et~al.}(2005)Arduini, Minnis, Smith~Jr, Ayers, Khaiyer, \&
  Heck}]{arduini2005sensitivity}
Arduini, R., Minnis, P., Smith~Jr, W., {et~al.} 2005, Sensitivity of
  satellite-retrieved cloud properties to the effective variance of cloud
  droplet size distribution, Tech. rep., Science Applications International
  Corporation, Hampton, Virginia; NASA~…

\bibitem[{{Beatty} {et~al.}(2019){Beatty}, {Marley}, {Gaudi}, {Col{\'o}n},
  {Fortney}, \& {Showman}}]{beatty19}
{Beatty}, T.~G., {Marley}, M.~S., {Gaudi}, B.~S., {et~al.} 2019, \aj, 158, 166,
  \dodoi{10.3847/1538-3881/ab33fc}

\bibitem[{{Bell} \& {Cowan}(2018)}]{bell18}
{Bell}, T.~J., \& {Cowan}, N.~B. 2018, \apjl, 857, L20,
  \dodoi{10.3847/2041-8213/aabcc8}

\bibitem[{{Bell} {et~al.}(2020){Bell}, {Dang}, {Cowan}, {Bean}, {D{\'e}sert},
  {Fortney}, {Keating}, {Kempton}, {Kreidberg}, {Line}, {Mansfield},
  {Parmentier}, {Stevenson}, {Swain}, \& {Zellem}}]{Bell20}
{Bell}, T.~J., {Dang}, L., {Cowan}, N.~B., {et~al.} 2020, arXiv e-prints,
  arXiv:2010.00687.
\newblock \doarXiv{2010.00687}

\bibitem[{Burrows \& Sharp(1999)}]{burrows1999chemical}
Burrows, A., \& Sharp, C. 1999, The Astrophysical Journal, 512, 843

\bibitem[{{Cowan} \& {Agol}(2011)}]{CowanAlgol2011}
{Cowan}, N.~B., \& {Agol}, E. 2011, \apj, 729, 54,
  \dodoi{10.1088/0004-637X/729/1/54}

\bibitem[{{de Rooij} \& {van der Stap}(1984)}]{de-rooij+1984}
{de Rooij}, W.~A., \& {van der Stap}, C.~C.~A.~H. 1984, \aap, 131, 237

\bibitem[{{Demory} {et~al.}(2013){Demory}, {de Wit}, {Lewis}, {Fortney},
  {Zsom}, {Seager}, {Knutson}, {Heng}, {Madhusudhan}, {Gillon}, {Barclay},
  {Desert}, {Parmentier}, \& {Cowan}}]{demory+2013}
{Demory}, B.-O., {de Wit}, J., {Lewis}, N., {et~al.} 2013, \apjl, 776, L25,
  \dodoi{10.1088/2041-8205/776/2/L25}

\bibitem[{{Dobbs-Dixon} \& {Lin}(2008)}]{dobbsdixon08}
{Dobbs-Dixon}, I., \& {Lin}, D.~N.~C. 2008, \apj, 673, 513,
  \dodoi{10.1086/523786}

\bibitem[{{Ehrenreich} {et~al.}(2020){Ehrenreich}, {Lovis}, {Allart}, {Zapatero
  Osorio}, {Pepe}, {Cristiani}, {Rebolo}, {Santos}, {Borsa}, {Demangeon},
  {Dumusque}, {Gonz{\'a}lez Hern{\'a}ndez}, {Casasayas-Barris},
  {S{\'e}gransan}, {Sousa}, {Abreu}, {Adibekyan}, {Affolter}, {Allende Prieto},
  {Alibert}, {Aliverti}, {Alves}, {Amate}, {Avila}, {Baldini}, {Bandy}, {Benz},
  {Bianco}, {Bolmont}, {Bouchy}, {Bourrier}, {Broeg}, {Cabral}, {Calderone},
  {Pall{\'e}}, {Cegla}, {Cirami}, {Coelho}, {Conconi}, {Coretti}, {Cumani},
  {Cupani}, {Dekker}, {Delabre}, {Deiries}, {D'Odorico}, {Di Marcantonio},
  {Figueira}, {Fragoso}, {Genolet}, {Genoni}, {G{\'e}nova Santos}, {Hara},
  {Hughes}, {Iwert}, {Kerber}, {Knudstrup}, {Land oni}, {Lavie}, {Lizon},
  {Lendl}, {Lo Curto}, {Maire}, {Manescau}, {Martins}, {M{\'e}gevand },
  {Mehner}, {Micela}, {Modigliani}, {Molaro}, {Monteiro}, {Monteiro},
  {Moschetti}, {M{\"u}ller}, {Nunes}, {Oggioni}, {Oliveira}, {Pariani},
  {Pasquini}, {Poretti}, {Rasilla}, {Redaelli}, {Riva}, {Santana Tschudi},
  {Santin}, {Santos}, {Segovia Milla}, {Seidel}, {Sosnowska}, {Sozzetti},
  {Span{\`o}}, {Su{\'a}rez Mascare{\~n}o}, {Tabernero}, {Tenegi}, {Udry},
  {Zanutta}, \& {Zerbi}}]{ehrenreich20}
{Ehrenreich}, D., {Lovis}, C., {Allart}, R., {et~al.} 2020, \nat, 580, 597,
  \dodoi{10.1038/s41586-020-2107-1}

\bibitem[{{Esteves} {et~al.}(2015){Esteves}, {De Mooij}, \&
  {Jayawardhana}}]{esteves+2015}
{Esteves}, L.~J., {De Mooij}, E. J.~W., \& {Jayawardhana}, R. 2015, \apj, 804,
  150, \dodoi{10.1088/0004-637X/804/2/150}

\bibitem[{{Gandhi} \& {Madhusudhan}(2019)}]{gandhi19}
{Gandhi}, S., \& {Madhusudhan}, N. 2019, \mnras, 485, 5817,
  \dodoi{10.1093/mnras/stz751}

\bibitem[{Gao {et~al.}(2017)Gao, Marley, Zahnle, Robinson, \&
  Lewis}]{gao2017sulfur}
Gao, P., Marley, M.~S., Zahnle, K., Robinson, T.~D., \& Lewis, N.~K. 2017, The
  Astronomical Journal, 153, 139

\bibitem[{Gao {et~al.}(2020)Gao, Thorngren, Lee, Fortney, Morley, Wakeford,
  Powell, Stevenson, \& Zhang}]{gao2020aerosol}
Gao, P., Thorngren, D.~P., Lee, G.~K., {et~al.} 2020, Nature Astronomy, 1

\bibitem[{{Garcia Munoz} \& {Isaak}(2015)}]{munoz+2015}
{Garcia Munoz}, A., \& {Isaak}, K.~G. 2015, Proceedings of the National Academy
  of Science, 112, 13461, \dodoi{10.1073/pnas.1509135112}

\bibitem[{Gaudi {et~al.}(2017)Gaudi, Stassun, Collins, Beatty, Zhou, Latham,
  Bieryla, Eastman, Siverd, Crepp, {et~al.}}]{gaudi2017giant}
Gaudi, B.~S., Stassun, K.~G., Collins, K.~A., {et~al.} 2017, Nature, 546, 514

\bibitem[{Guillot(2010)}]{guillot2010radiative}
Guillot, T. 2010, Astronomy \& Astrophysics, 520, A27

\bibitem[{{Harada} {et~al.}(2019){Harada}, {Kempton}, {Rauscher}, {Roman}, \&
  {Brinjikji}}]{harada20}
{Harada}, C.~K., {Kempton}, E. M.~R., {Rauscher}, E., {Roman}, M., \&
  {Brinjikji}, M. 2019, arXiv e-prints, arXiv:1912.02268.
\newblock \doarXiv{1912.02268}

\bibitem[{{Helling} {et~al.}(2019){Helling}, {Iro}, {Corrales}, {Samra},
  {Ohno}, {Alam}, {Steinrueck}, {Lew}, {Molaverdikhani}, {MacDonald},
  {Herbort}, {Woitke}, \& {Parmentier}}]{helling19}
{Helling}, C., {Iro}, N., {Corrales}, L., {et~al.} 2019, \aap, 631, A79,
  \dodoi{10.1051/0004-6361/201935771}

\bibitem[{{Heng} {et~al.}(2011){Heng}, {Menou}, \& {Phillipps}}]{Heng2011}
{Heng}, K., {Menou}, K., \& {Phillipps}, P.~J. 2011, \mnras, 413, 2380,
  \dodoi{10.1111/j.1365-2966.2011.18315.x}

\bibitem[{Iro {et~al.}(2005)Iro, Bezard, \& Guillot}]{iro2005timeconstants}
Iro, N., Bezard, B., \& Guillot, T. 2005, Astronomy \& Astrophysics, 436, 719

\bibitem[{Johnson \& Christy(1974)}]{johnson1974optical}
Johnson, P., \& Christy, R. 1974, Physical review B, 9, 5056

\bibitem[{{Kataria} {et~al.}(2015){Kataria}, {Showman}, {Fortney}, {Stevenson},
  {Line}, {Kreidberg}, {Bean}, \& {D{\'e}sert}}]{Kataria2015}
{Kataria}, T., {Showman}, A.~P., {Fortney}, J.~J., {et~al.} 2015, \apj, 801,
  86, \dodoi{10.1088/0004-637X/801/2/86}

\bibitem[{{Kataria} {et~al.}(2016){Kataria}, {Sing}, {Lewis}, {Visscher},
  {Showman}, {Fortney}, \& {Marley}}]{kataria+2016}
{Kataria}, T., {Sing}, D.~K., {Lewis}, N.~K., {et~al.} 2016, \apj, 821, 9,
  \dodoi{10.3847/0004-637X/821/1/9}

\bibitem[{{Keating} {et~al.}(2019){Keating}, {Cowan}, \& {Dang}}]{keating+2019}
{Keating}, D., {Cowan}, N.~B., \& {Dang}, L. 2019, Nature Astronomy, 426,
  \dodoi{10.1038/s41550-019-0859-z}

\bibitem[{{Keating} {et~al.}(2020){Keating}, {Stevenson}, {Cowan}, {Rauscher},
  {Bean}, {Bell}, {Dang}, {Deming}, {D{\'e}sert}, {Feng}, {Fortney}, {Kataria},
  {Kempton}, {Lewis}, {Line}, {Mansfield}, {May}, {Morley}, \&
  {Showman}}]{keating2020}
{Keating}, D., {Stevenson}, K.~B., {Cowan}, N.~B., {et~al.} 2020, arXiv
  e-prints, arXiv:2004.00014.
\newblock \doarXiv{2004.00014}

\bibitem[{{Kitzmann} \& {Heng}(2018)}]{kitzmann+2018}
{Kitzmann}, D., \& {Heng}, K. 2018, \mnras, 475, 94,
  \dodoi{10.1093/mnras/stx3141}

\bibitem[{{Komacek} \& {Showman}(2016)}]{KomacekShowman2016}
{Komacek}, T.~D., \& {Showman}, A.~P. 2016, \apj, 821, 16,
  \dodoi{10.3847/0004-637X/821/1/16}

\bibitem[{{Komacek} {et~al.}(2019){Komacek}, {Showman}, \&
  {Parmentier}}]{komacek19}
{Komacek}, T.~D., {Showman}, A.~P., \& {Parmentier}, V. 2019, \apj, 881, 152,
  \dodoi{10.3847/1538-4357/ab338b}

\bibitem[{{Komacek} \& {Tan}(2018)}]{komacek18}
{Komacek}, T.~D., \& {Tan}, X. 2018, Research Notes of the American
  Astronomical Society, 2, 36, \dodoi{10.3847/2515-5172/aac5e7}

\bibitem[{{Langton} \& {Laughlin}(2007)}]{langton07}
{Langton}, J., \& {Laughlin}, G. 2007, \apjl, 657, L113, \dodoi{10.1086/513185}

\bibitem[{{Latham} {et~al.}(2010){Latham}, {Borucki}, {Koch}, {Brown},
  {Buchhave}, {Basri}, {Batalha}, {Caldwell}, {Cochran}, {Dunham}, {F{\H
  u}r{\'e}sz}, {Gautier}, {Geary}, {Gilliland}, {Howell}, {Jenkins},
  {Lissauer}, {Marcy}, {Monet}, {Rowe}, \& {Sasselov}}]{latham+2010}
{Latham}, D.~W., {Borucki}, W.~J., {Koch}, D.~G., {et~al.} 2010, \apjl, 713,
  L140, \dodoi{10.1088/2041-8205/713/2/L140}

\bibitem[{{Lee} {et~al.}(2016){Lee}, {Dobbs-Dixon}, {Helling}, {Bognar}, \&
  {Woitke}}]{lee+2016}
{Lee}, G., {Dobbs-Dixon}, I., {Helling}, C., {Bognar}, K., \& {Woitke}, P.
  2016, \aap, 594, A48, \dodoi{10.1051/0004-6361/201628606}

\bibitem[{{Lee} {et~al.}(2015){Lee}, {Helling}, {Dobbs-Dixon}, \&
  {Juncher}}]{lee15}
{Lee}, G., {Helling}, C., {Dobbs-Dixon}, I., \& {Juncher}, D. 2015, \aap, 580,
  A12, \dodoi{10.1051/0004-6361/201525982}

\bibitem[{{Lee} {et~al.}(2017){Lee}, {Wood}, {Dobbs-Dixon}, {Rice}, \&
  {Helling}}]{lee+2017}
{Lee}, G.~K.~H., {Wood}, K., {Dobbs-Dixon}, I., {Rice}, A., \& {Helling}, C.
  2017, \aap, 601, A22, \dodoi{10.1051/0004-6361/201629804}

\bibitem[{{Lines} {et~al.}(2019){Lines}, {Mayne}, {Manners}, {Boutle},
  {Drummond}, {Mikal-Evans}, {Kohary}, \& {Sing}}]{Lines+2019}
{Lines}, S., {Mayne}, N.~J., {Manners}, J., {et~al.} 2019, \mnras, 488, 1332,
  \dodoi{10.1093/mnras/stz1788}

\bibitem[{{Lines} {et~al.}(2018){Lines}, {Mayne}, {Boutle}, {Manners}, {Lee},
  {Helling}, {Drummond}, {Amundsen}, {Goyal}, {Acreman}, {Tremblin}, \&
  {Kerslake}}]{lines+2018}
{Lines}, S., {Mayne}, N.~J., {Boutle}, I.~A., {et~al.} 2018, \aap, 615, A97,
  \dodoi{10.1051/0004-6361/201732278}

\bibitem[{L{\'o}pez(1977)}]{lopez_1977lognormal}
L{\'o}pez, R.~E. 1977, Monthly Weather Review, 105, 865

\bibitem[{{Lothringer} {et~al.}(2018){Lothringer}, {Barman}, \&
  {Koskinen}}]{lothringer18}
{Lothringer}, J.~D., {Barman}, T., \& {Koskinen}, T. 2018, \apj, 866, 27,
  \dodoi{10.3847/1538-4357/aadd9e}

\bibitem[{Lunine {et~al.}(1989)Lunine, Hubbard, Burrows, Wang, \&
  Garlow}]{lunine1989effect}
Lunine, J.~I., Hubbard, W., Burrows, A., Wang, Y.-P., \& Garlow, K. 1989, The
  Astrophysical Journal, 338, 314

\bibitem[{Lunine \& Hunten(1989)}]{lunine1989abundance}
Lunine, J.~I., \& Hunten, D.~M. 1989, Planetary and space science, 37, 151

\bibitem[{{Mansfield} {et~al.}(2020){Mansfield}, {Bean}, {Stevenson},
  {Komacek}, {Bell}, {Tan}, {Malik}, {Beatty}, {Wong}, {Cowan}, {Dang},
  {D{\'e}sert}, {Fortney}, {Gaudi}, {Keating}, {Kempton}, {Kreidberg}, {Line},
  {Parmentier}, {Stassun}, {Swain}, \& {Zellem}}]{mansfield20}
{Mansfield}, M., {Bean}, J.~L., {Stevenson}, K.~B., {et~al.} 2020, \apjl, 888,
  L15, \dodoi{10.3847/2041-8213/ab5b09}

\bibitem[{{Marley} {et~al.}(2013){Marley}, {Ackerman}, {Cuzzi}, \&
  {Kitzmann}}]{Marley2013exoclouds}
{Marley}, M.~S., {Ackerman}, A.~S., {Cuzzi}, J.~N., \& {Kitzmann}, D. 2013,
  {Clouds and Hazes in Exoplanet Atmospheres} (University of Arizona Press),
  367--391, \dodoi{10.2458/azu_uapress_9780816530595-ch15}

\bibitem[{{Mbarek} \& {Kempton}(2016)}]{Mbarek&Kempton2016}
{Mbarek}, R., \& {Kempton}, E.~M.-R. 2016, \apj, 827, 121,
  \dodoi{10.3847/0004-637X/827/2/121}

\bibitem[{{Mendon{\c c}a} {et~al.}(2018){Mendon{\c c}a}, {Malik}, {Demory}, \&
  {Heng}}]{Mendoca2018}
{Mendon{\c c}a}, J.~M., {Malik}, M., {Demory}, B.-O., \& {Heng}, K. 2018, \aj,
  155, 150, \dodoi{10.3847/1538-3881/aaaebc}

\bibitem[{{Mishchenko} {et~al.}(1999){Mishchenko}, {Dlugach}, {Yanovitskij}, \&
  {Zakharova}}]{mishchenko+1999}
{Mishchenko}, M.~I., {Dlugach}, Z.~M., {Yanovitskij}, E.~G., \& {Zakharova},
  N.~T. 1999, \jqsrt, 63, 409, \dodoi{10.1016/S0022-4073(99)00028-X}

\bibitem[{{Moses} {et~al.}(2013){Moses}, {Madhusudhan}, {Visscher}, \&
  {Freedman}}]{Moses2013}
{Moses}, J.~I., {Madhusudhan}, N., {Visscher}, C., \& {Freedman}, R.~S. 2013,
  \apj, 763, 25, \dodoi{10.1088/0004-637X/763/1/25}

\bibitem[{{Oreshenko} {et~al.}(2016{\natexlab{a}}){Oreshenko}, {Heng}, \&
  {Demory}}]{Oreshenko2015}
{Oreshenko}, M., {Heng}, K., \& {Demory}, B.-O. 2016{\natexlab{a}}, \mnras,
  457, 3420, \dodoi{10.1093/mnras/stw133}

\bibitem[{{Oreshenko} {et~al.}(2016{\natexlab{b}}){Oreshenko}, {Heng}, \&
  {Demory}}]{oreshenko+2016}
---. 2016{\natexlab{b}}, \mnras, 457, 3420, \dodoi{10.1093/mnras/stw133}

\bibitem[{{Parmentier} {et~al.}(2016{\natexlab{a}}){Parmentier}, {Fortney},
  {Showman}, {Morley}, \& {Marley}}]{parmentier+2016}
{Parmentier}, V., {Fortney}, J.~J., {Showman}, A.~P., {Morley}, C., \&
  {Marley}, M.~S. 2016{\natexlab{a}}, \apj, 828, 22,
  \dodoi{10.3847/0004-637X/828/1/22}

\bibitem[{{Parmentier} {et~al.}(2016{\natexlab{b}}){Parmentier}, {Fortney},
  {Showman}, {Morley}, \& {Marley}}]{Parmentier2016}
---. 2016{\natexlab{b}}, \apj, 828, 22, \dodoi{10.3847/0004-637X/828/1/22}

\bibitem[{{Parmentier} {et~al.}(2013){Parmentier}, {Showman}, \&
  {Lian}}]{Parmentier2013}
{Parmentier}, V., {Showman}, A.~P., \& {Lian}, Y. 2013, \aap, 558, A91,
  \dodoi{10.1051/0004-6361/201321132}

\bibitem[{{Parmentier} {et~al.}(2018){Parmentier}, {Line}, {Bean}, {Mansfield},
  {Kreidberg}, {Lupu}, {Visscher}, {D{\'e}sert}, {Fortney}, {Deleuil},
  {Arcangeli}, {Showman}, \& {Marley}}]{parmentier18}
{Parmentier}, V., {Line}, M.~R., {Bean}, J.~L., {et~al.} 2018, \aap, 617, A110,
  \dodoi{10.1051/0004-6361/201833059}

\bibitem[{{Perez-Becker} \& {Showman}(2013)}]{PerezBeckerShowman2013}
{Perez-Becker}, D., \& {Showman}, A.~P. 2013, \apj, 776, 134,
  \dodoi{10.1088/0004-637X/776/2/134}

\bibitem[{{Perna} {et~al.}(2012){Perna}, {Heng}, \& {Pont}}]{perna12}
{Perna}, R., {Heng}, K., \& {Pont}, F. 2012, \apj, 751, 59,
  \dodoi{10.1088/0004-637X/751/1/59}

\bibitem[{{Perna} {et~al.}(2010){Perna}, {Menou}, \& {Rauscher}}]{Perna2010}
{Perna}, R., {Menou}, K., \& {Rauscher}, E. 2010, \apj, 719, 1421,
  \dodoi{10.1088/0004-637X/719/2/1421}

\bibitem[{{Powell} {et~al.}(2019){Powell}, {Louden}, {Kreidberg}, {Zhang},
  {Gao}, \& {Parmentier}}]{powell19}
{Powell}, D., {Louden}, T., {Kreidberg}, L., {et~al.} 2019, \apj, 887, 170,
  \dodoi{10.3847/1538-4357/ab55d9}

\bibitem[{{Powell} {et~al.}(2018){Powell}, {Zhang}, {Gao}, \&
  {Parmentier}}]{powell+2018}
{Powell}, D., {Zhang}, X., {Gao}, P., \& {Parmentier}, V. 2018, \apj, 860, 18,
  \dodoi{10.3847/1538-4357/aac215}

\bibitem[{{Rauscher} \& {Menou}(2010)}]{rauscher+2010}
{Rauscher}, E., \& {Menou}, K. 2010, \apj, 714, 1334,
  \dodoi{10.1088/0004-637X/714/2/1334}

\bibitem[{{Rauscher} \& {Menou}(2012)}]{rauscher+2012}
---. 2012, \apj, 750, 96, \dodoi{10.1088/0004-637X/750/2/96}

\bibitem[{{Rauscher} \& {Menou}(2013)}]{rauscher13}
---. 2013, \apj, 764, 103, \dodoi{10.1088/0004-637X/764/1/103}

\bibitem[{{Rogers} \& {Komacek}(2014)}]{Rogers2014}
{Rogers}, T.~M., \& {Komacek}, T.~D. 2014, \apj, 794, 132,
  \dodoi{10.1088/0004-637X/794/2/132}

\bibitem[{{Roman} \& {Rauscher}(2017)}]{Roman&Rauscher2017}
{Roman}, M., \& {Rauscher}, E. 2017, \apj, 850, 17,
  \dodoi{10.3847/1538-4357/aa8ee4}

\bibitem[{{Roman} \& {Rauscher}(2019)}]{Roman&Rauscher2019}
---. 2019, \apj, 872, 1, \dodoi{10.3847/1538-4357/aafdb5}

\bibitem[{Roman {et~al.}(2013)Roman, Banfield, \& Gierasch}]{roman2013saturn}
Roman, M.~T., Banfield, D., \& Gierasch, P.~J. 2013, Icarus, 225, 93

\bibitem[{{Rossow}(1978)}]{Rossow1978}
{Rossow}, W.~B. 1978, \icarus, 36, 1, \dodoi{10.1016/0019-1035(78)90072-6}

\bibitem[{Seager {et~al.}(2005)Seager, Richardson, Hansen, Menou, Cho, \&
  Deming}]{seager2005dayside}
Seager, S., Richardson, L., Hansen, B., {et~al.} 2005, The Astrophysical
  Journal, 632, 1122

\bibitem[{Shannon {et~al.}(2017)Shannon, Lafuente, Shannon, Downs, \&
  Fischer}]{shannon2017refractive}
Shannon, R.~C., Lafuente, B., Shannon, R.~D., Downs, R.~T., \& Fischer, R.~X.
  2017, American Mineralogist: Journal of Earth and Planetary Materials, 102,
  1906

\bibitem[{{Showman} \& {Guillot}(2002)}]{Showman&Guillot2002}
{Showman}, A.~P., \& {Guillot}, T. 2002, \aap, 385, 166,
  \dodoi{10.1051/0004-6361:20020101}

\bibitem[{{Showman} {et~al.}(2015){Showman}, {Lewis}, \& {Fortney}}]{showman15}
{Showman}, A.~P., {Lewis}, N.~K., \& {Fortney}, J.~J. 2015, \apj, 801, 95,
  \dodoi{10.1088/0004-637X/801/2/95}

\bibitem[{{Showman} \& {Polvani}(2011)}]{ShowmanPolvani2011}
{Showman}, A.~P., \& {Polvani}, L.~M. 2011, \apj, 738, 71,
  \dodoi{10.1088/0004-637X/738/1/71}

\bibitem[{{Stevenson} {et~al.}(2014){Stevenson}, {D{\'e}sert}, {Line}, {Bean},
  {Fortney}, {Showman}, {Kataria}, {Kreidberg}, {McCullough}, {Henry},
  {Charbonneau}, {Burrows}, {Seager}, {Madhusudhan}, {Williamson}, \&
  {Homeier}}]{stevenson+2014}
{Stevenson}, K.~B., {D{\'e}sert}, J.-M., {Line}, M.~R., {et~al.} 2014, Science,
  346, 838, \dodoi{10.1126/science.1256758}

\bibitem[{{Stevenson} {et~al.}(2017){Stevenson}, {Line}, {Bean}, {D{\'e}sert},
  {Fortney}, {Showman}, {Kataria}, {Kreidberg}, \& {Feng}}]{stevenson+2017}
{Stevenson}, K.~B., {Line}, M.~R., {Bean}, J.~L., {et~al.} 2017, \aj, 153, 68,
  \dodoi{10.3847/1538-3881/153/2/68}

\bibitem[{{Tan} \& {Komacek}(2019)}]{tan19}
{Tan}, X., \& {Komacek}, T.~D. 2019, \apj, 886, 26,
  \dodoi{10.3847/1538-4357/ab4a76}

\bibitem[{{Thorngren} {et~al.}(2019){Thorngren}, {Gao}, \&
  {Fortney}}]{thorngren+2019}
{Thorngren}, D., {Gao}, P., \& {Fortney}, J.~J. 2019, \apjl, 884, L6,
  \dodoi{10.3847/2041-8213/ab43d0}

\bibitem[{{Toon} {et~al.}(1989){Toon}, {McKay}, {Ackerman}, \&
  {Santhanam}}]{toon+1989}
{Toon}, O.~B., {McKay}, C.~P., {Ackerman}, T.~P., \& {Santhanam}, K. 1989,
  \jgr, 94, 16287, \dodoi{10.1029/JD094iD13p16287}

\bibitem[{Wan {et~al.}(2019)Wan, Zhang, Woolf, Hessel, Rensberg, Hensley, Xiao,
  Shahsafi, Salman, Richter, {et~al.}}]{wan2019optical}
Wan, C., Zhang, Z., Woolf, D., {et~al.} 2019, Annalen der Physik, 531, 1900188

\bibitem[{{Zhang} {et~al.}(2018){Zhang}, {Knutson}, {Kataria}, {Schwartz},
  {Cowan}, {Showman}, {Burrows}, {Fortney}, {Todorov}, {Desert}, {Agol}, \&
  {Deming}}]{zhang18}
{Zhang}, M., {Knutson}, H.~A., {Kataria}, T., {et~al.} 2018, \aj, 155, 83,
  \dodoi{10.3847/1538-3881/aaa458}

\end{thebibliography}
\bibliographystyle{aasjournal}

\appendix

\section{Cloud Scattering Properties} 
\label{sec:appendix}
\begin{figure*}[b!]
\includegraphics[clip, trim=.0in 0in 0in 0in,width=\textwidth]{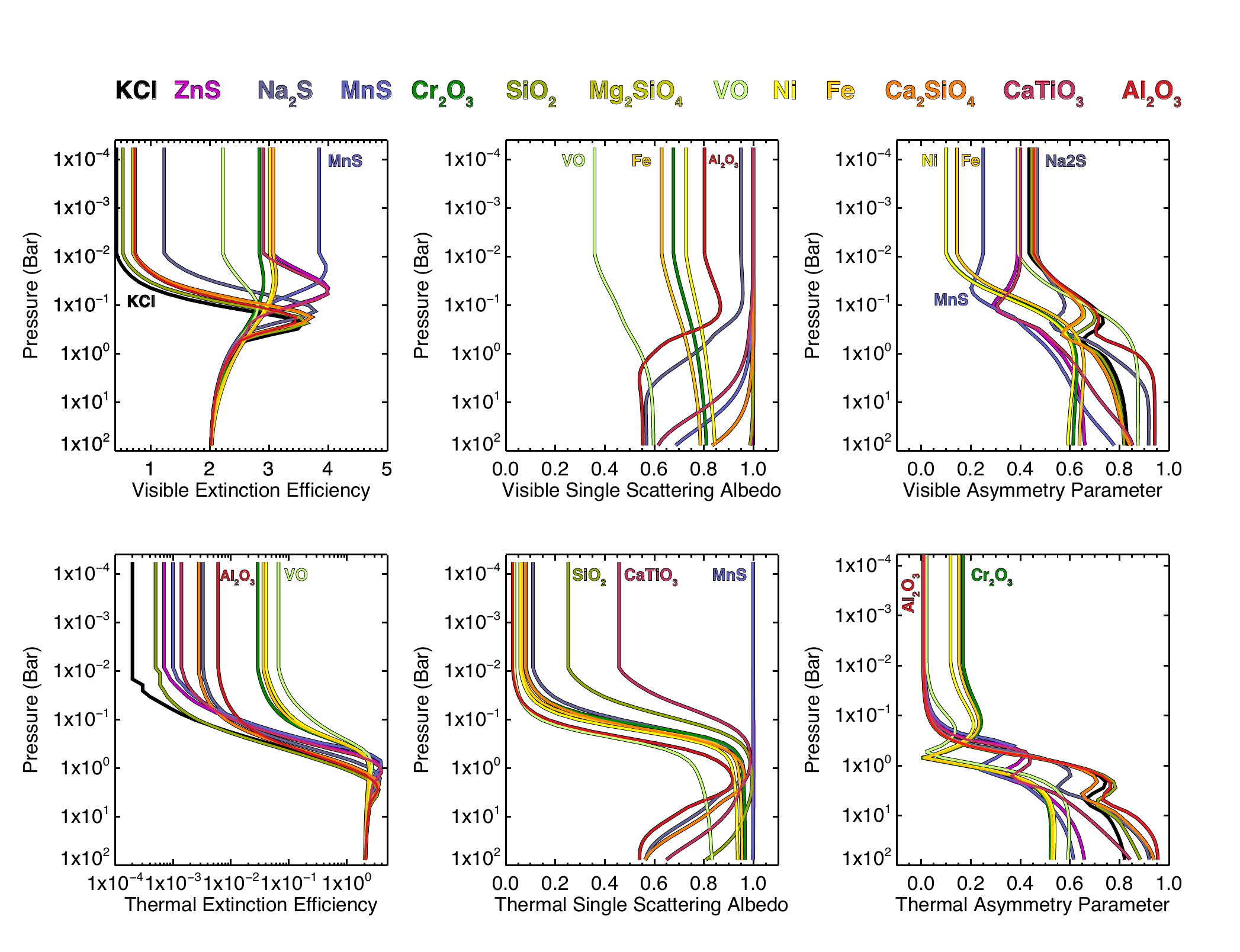}
\caption{Scattering parameters used for each of the thirteen cloud compounds assumed in the models as a function of pressure. The plot colors correspond to the compounds listed above, with several curves additionally labeled to facilitate identification.  The extinction efficiency, single scattering albedo, and asymmetry parameters (left to right) were computed for both the visible (top) and thermal (bottom) channels, evaluated at 650 nm and 5 $\mu$m, respectively \citep[as in][]{Roman&Rauscher2019}.  The vertical dependence comes from the assumed particle size gradient. Small particles near the top of the model yield a wide range of albedos in both the visible and thermal channels.  The extinction also shows a significant range of values, although the magnitude of thermal extinction is consistently much less than the visible extinction.  Asymmetry parameters generally increase with pressure (i.e.\ particle size) for all compounds, producing fairly isotropic scattering in the thermal at IR photospheric pressures and strongly asymmetric scattering at higher pressures.}
\label{fig:scatparams}
\end{figure*}

We compute scattering parameters using Mie theory, the indices of refraction listed in Table \ref{table:clouds}, and the assumed particle sizes and distribution (see Figure \ref{fig:scatparams}). These include: the extinction efficiency, which is used to convert the cloud masses into optical thicknesses; the single scattering albedo, which describes the extent to which a cloud absorbs or scatters radiation; and the asymmetry parameter, which describes the directionality of scatterers. These parameters are varied in height due to their dependence on the assumed particle size.  We assume a vertical gradient in particle radii roughly based on the vertical gradient found in physical models \citep{Parmentier2013,Lines+2019}; particle radii range from 0.1 $\mu$m at pressures $<$0.01 bar and increase exponentially in radius to nearly 80 $\mu$m at the base of the model. 

The range of the parameters shown in Figure \ref{fig:scatparams} displays the sensitivity to both the particle size and composition. The general trend of increasing scattering asymmetry with increasing pressure seen in both wavelength channels is a consequence of the relatively robust trend of smaller particles (relative to the wavelength) to scatter more isotropically. Likewise, the extinctions at depth converge towards the theoretical value of $Q_e = 2$ (due to geometric cross section plus diffraction) when particle size is much greater than the radiation's wavelength. The variation beyond this is due to the range in refractive indices. For example, the low imaginary indices of refraction makes MnS an extremely conservative scatterer whereas other compounds tend to be more absorbing in the infrared; however, as a scatterer, its efficiency in the thermal is relatively low compared to other compounds, so its effect will be reduced, even though it has the greatest extinction efficiency in the visible. Additionally, MnS is a relatively minor species in abundance compared to the silicate and metal clouds, and its relatively low condensation temperature means that it condenses less broadly. Ultimately, it is the combination of all of these factors for each compound that together shape the heating rates and emission from the atmosphere.

\section{Temperature-Pressure Profiles} 
\label{sec:appendix_tp}
\begin{figure*}[p]
    \centering
    \includegraphics[clip, trim=0.3in 0.0in 0.in 0.0in,scale=0.81]{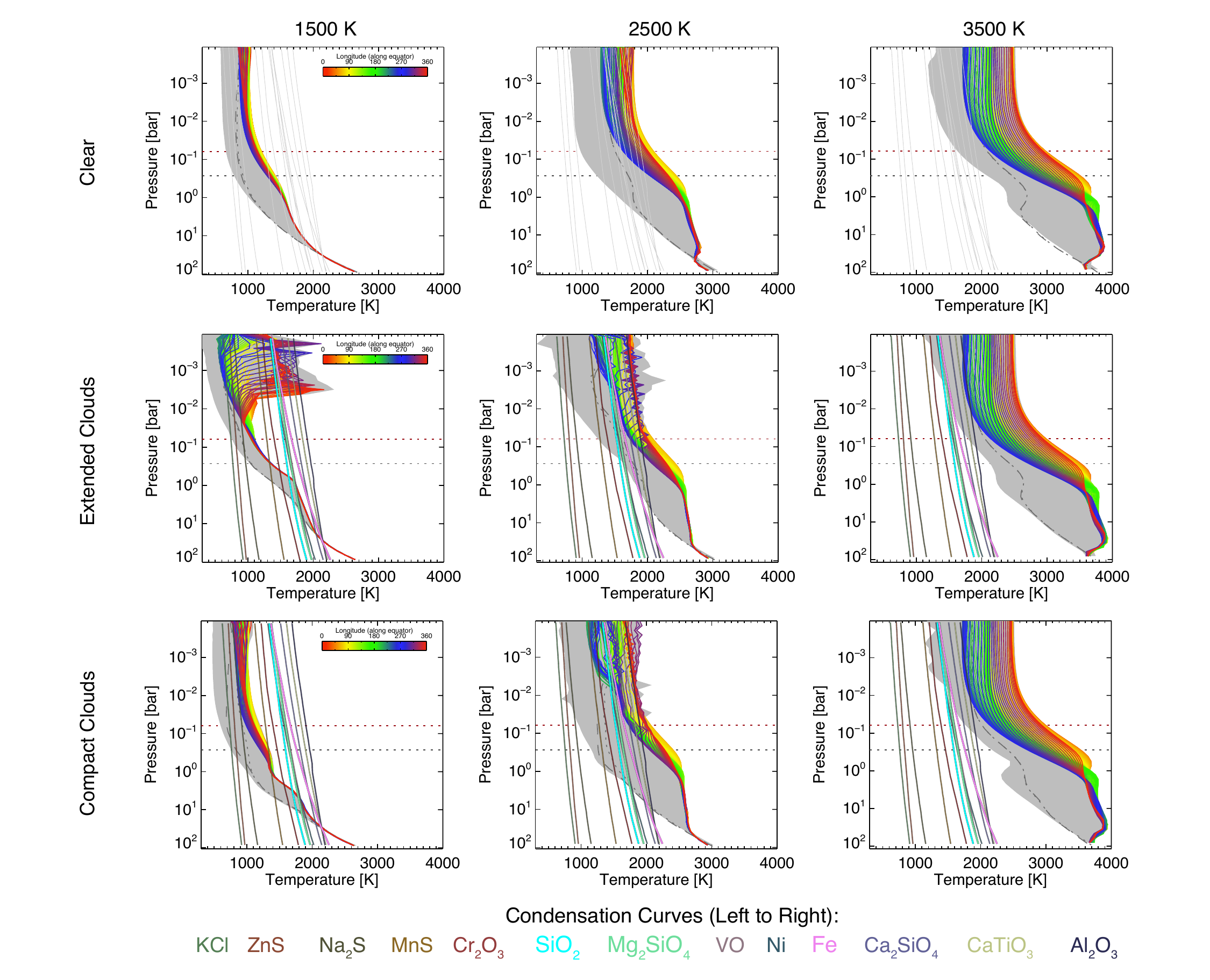}
    \caption{Temperature profiles from simulations at three different irradiation temperatures (1,500 K, 2,500 K, and 3,500 K) for our different cloud models --- clear, extended clouds, and compact clouds, indicated on the left, assuming 10$\%$ of the vapor mass condenses and a surface gravity of 10 m s$^{-2}$.  Temperature profiles along the equator are shown in colors corresponding to longitude; north and south polar temperature profiles are delineated in dotted and dashed lines, respectively; profiles for all other locations fall within the grey shaded envelope.  The horizontal dotted line marks the pressure where $\tau_{gas}$=2/3 in the visible (black) and thermal (red) measured from the top of the clear atmosphere. Condensation curves for each cloud species are superimposed in light grey in the clear cases and in color in the cloudy cases, corresponding to the species listed below. The simulations for clear atmospheres (top row) show that the temperatures smoothly increase in both value and range with increasing irradiation temperature.  When clouds are present (2nd \& 3rd rows), their bases form where temperature profiles cross the condensation curves, altering the profiles depending on the cloud mass.  In the colder cases, nearly all temperature profiles fall below the condensation curves, with the massive silicate and iron clouds condensing starting deep within the atmosphere; if vertically extended, clouds will cover nearly the entire planet, altering the temperature profiles through scattering and absorption.  As the irradiation temperature increases, profiles move farther to the right in the plots and cross fewer condensation curves, leading to fewer clouds in the hottest regions of the dayside. Finally, at the hottest irradiation temperatures, few profiles cross the condensation curves, resulting in only minor deviations from the clear cases at higher latitudes.}
    \label{fig:tprofs_vs_mods_a}
\end{figure*}

\begin{figure*}[ht]
    \centering
    \includegraphics[clip, trim=0.3in 0.0in 0.in 0.0in,scale=0.70]{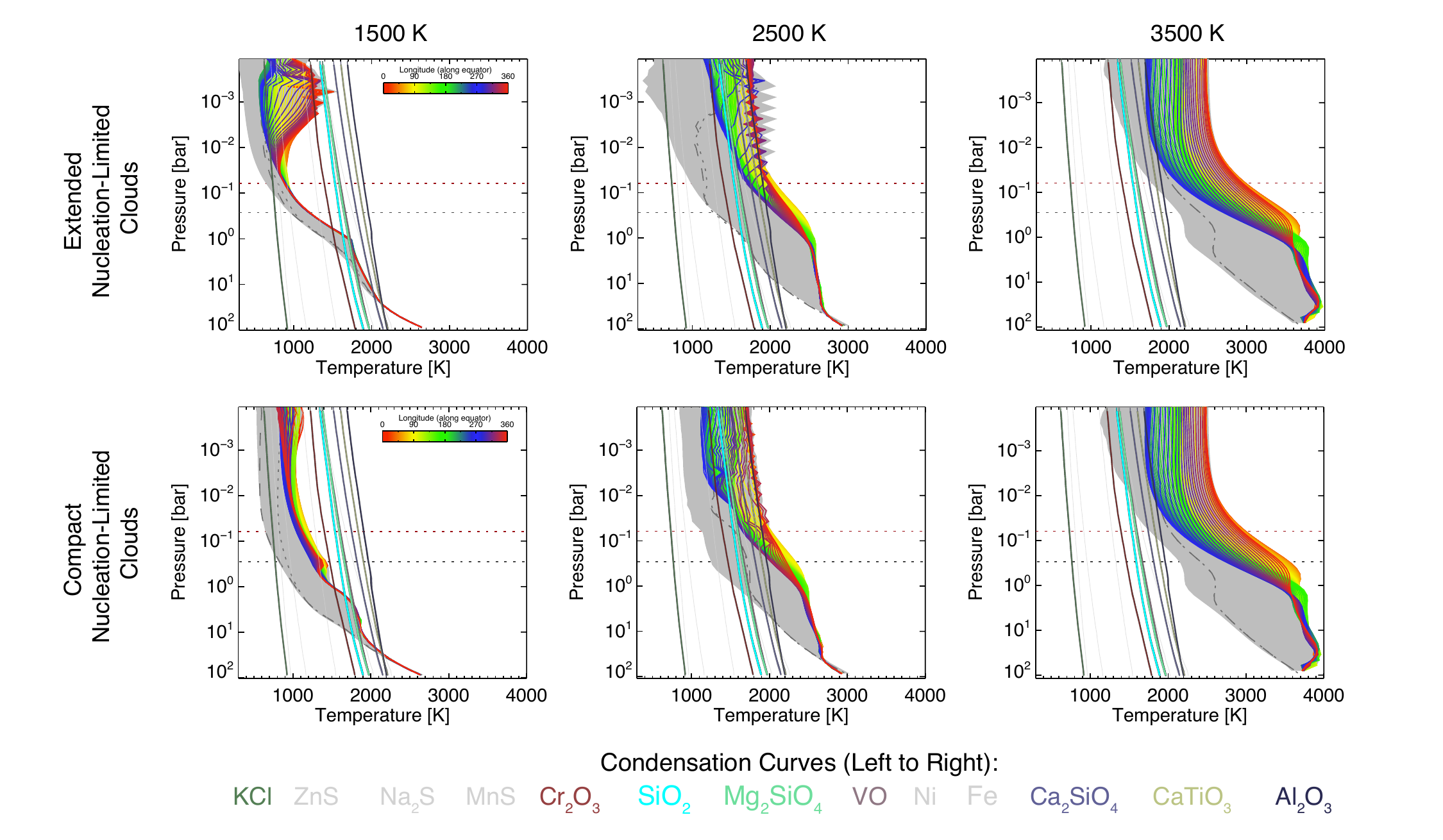}
    \caption{As in Figure \ref{fig:tprofs_vs_mods_a}, but now for models with nucleation-limited cloud models, in which ZnS, Na$_2$S, MnS, Fe, and Ni clouds are absent given their presumed lower nucleation rates. With Fe absent, stellar heating in the colder cases is more modest, resulting in smoother dayside profiles. Remaining Cr$_2$O$_3$ and Al$_2$O$_3$ clouds serve as moderate absorbers that modify the dayside temperature profiles.}
    \label{fig:tprofs_vs_mods_b}
\end{figure*}

Temperature profiles for a range of cases are shown in Figures \ref{fig:tprofs_vs_mods_a} and \ref{fig:tprofs_vs_mods_b}.  Careful inspection of the profiles in relation to the condensation curves reveals how different cloud species alter the atmospheric temperature due to their radiative effects.  Where temperature profiles and condensation curves intersect as a function of irradiation temperature is key to explaining the collective trends in our simulations and the dependence on different cloud modeling assumptions.  At colder temperatures, most of the clouds interesect the condensation curves at greater depths, shaping the profiles through thermal scattering.  Only in cases where the clouds are allowed to extend vertically above the clear visible photospheres do they warm the upper atmosphere through absorption and scattering of stellar radiation.  At intermediate temperatures, this intersection between profiles and condensation curves is pushed up to heights within the visible atmosphere, allowing for stellar absorption regardless of whether the clouds are compact or extended.   

The relative importance of the silicates (SiO$_2$ and Mg$_2$SiO$_4$) as scatters and Fe, Al$_2$O$_3$, and Cr$_2$O$_3$ as absorbers is evident from their sometimes dramatic effects on the vertical profiles. Extended cases that include iron clouds (Figure \ref{fig:tprofs_vs_mods_a}, second row) show the most severe warming, while the nucleation limited cases warm more modestly as they continue to cool the underlying atmosphere. Though less massive than Fe, the high temperature Al$_2$O$_3$ (corundum) clouds can have a controlling effect on the substellar temperature profiles at moderate irradiation temperatures through radiative feedback in our simple modeling scheme, as previously noted by \cite{Roman&Rauscher2019} and \cite{Lines+2019}. This can be seen in the substellar profiles at $T_{\mathrm{irr}} =$ 2,500 K, (middle column in Figures \ref{fig:tprofs_vs_mods_a} and \ref{fig:tprofs_vs_mods_b}), in which the temperature profiles align themselves with (or oscillate around) the condensation curves, forming a quasi-equilibrium between condensation and vaporization due to aerosol heating. This behavior can be found in our modeling whenever temperature profiles fall just below condensation curves for strongly absorbing species (Fe, Al$_2$O$_3$, and Cr$_2$O$_3$). \cite{Lines+2019} reported that such behavior was not found in their modeling of HD 209458b, but they noted the their temperatures generally fell far below the Al$_2$O$_3$ condensation point. The difference in behavior, however, may be due to \cite{Lines+2019} including particle sedimentation in their modeling, which effectively reduces the condensable vapor from the cloud layer and potentially breaks the cycle of vaporization and re-condensation if the vertical mixing is insufficient to homogenize the gaseous composition.  While the mechanism that produces this quasi-equilibrium is easily understood in our simple model, more advanced cloud modeling at similar temperatures will be needed to confidently determine if such behavior is unphysical or should be expected in reality. 

The effects of clouds are amplified when the cloud mass is increased, raising the thermal and visible photospheres to dramatically lower pressures and hotter dayside temperatures.  Conversely, when the gravity is increased, the lower mass per pressure interval increases the pressure of the photospheres, stretching the stellar heating to higher pressures with only marginal effects on the consequent observables. 
%\textcolor{red}{Possible figures:}

\clearpage

\begin{figure*}[hb]
    \centering
    \includegraphics[clip, trim=.3in 0.0in 0.in 0.0in,scale=0.70]{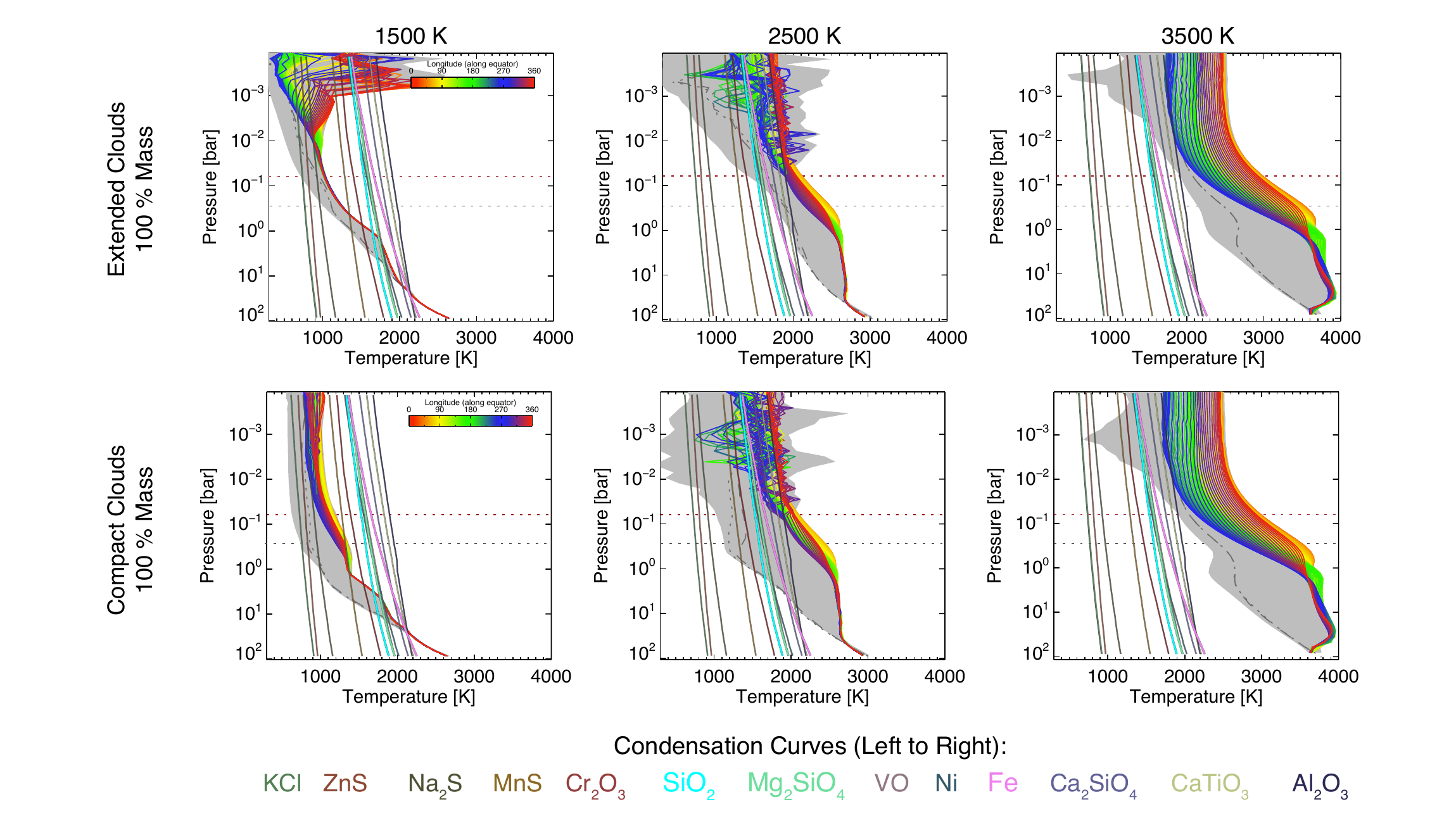}
    \caption{As in Figure \ref{fig:tprofs_vs_mods_a}, but now for models with 100$\%$ of the vapor mass allowed to condense. When clouds reach their theoretical mass upper-limit, their radiative effects are amplified, producing higher and hotter photospheres on the dayside and greater thermal scattering in the deep atmosphere. Even in the higher temperature cases, previously negligible clouds are shown to radiatively heat and cool at higher latitudes on the nightside.}
    \label{fig:tprofs_vs_mods_c}
\end{figure*}

\clearpage

\begin{figure*}
    \centering
    \includegraphics[clip, trim=0.3in 0.0in 0.in 0.0in,scale=0.70]{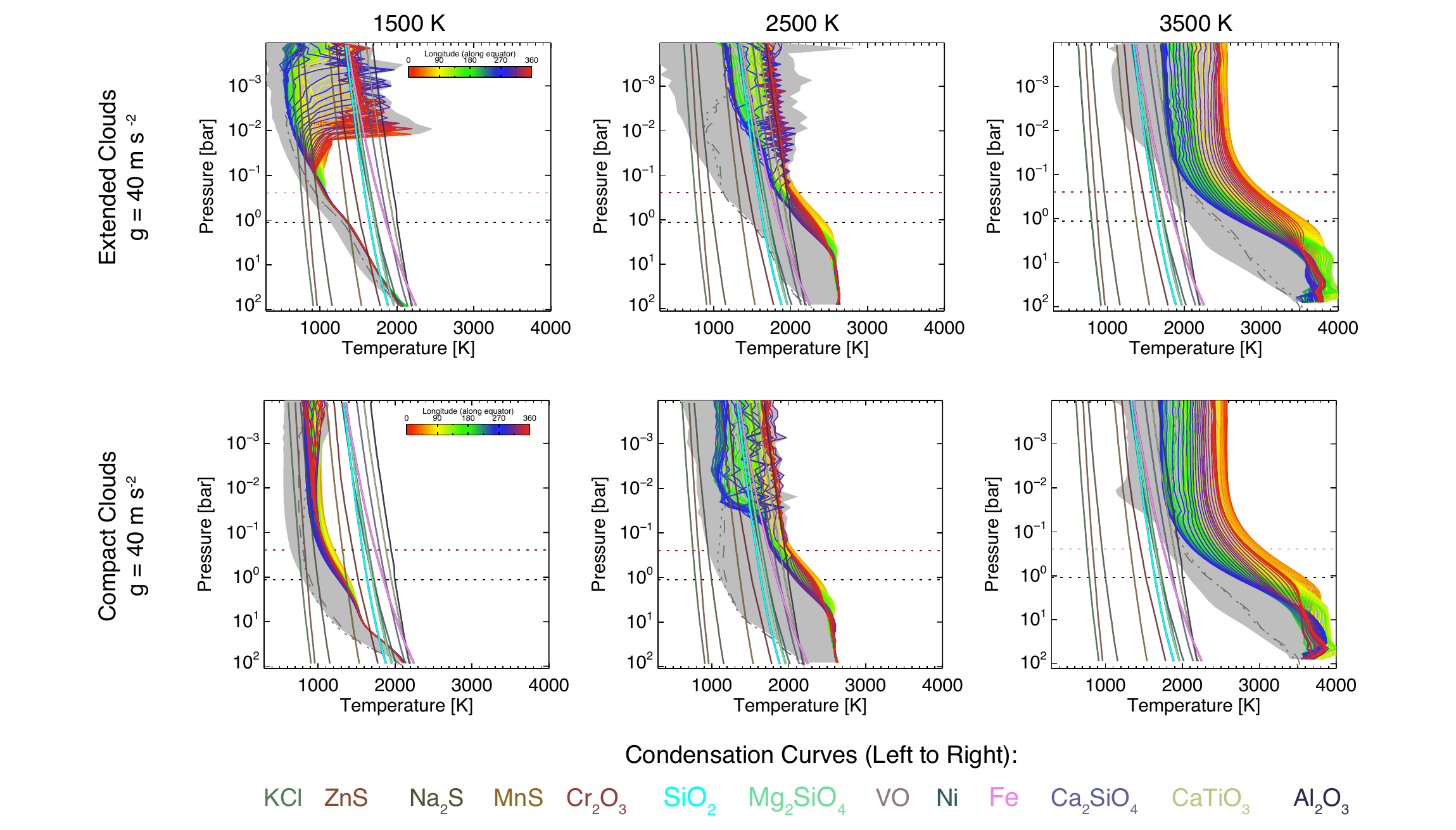}
    \caption{As in Figure \ref{fig:tprofs_vs_mods_a}, but now with a surface gravity of 40 m s$^2$ with 10$\%$ mass condensation. When the gravity is increased relative to the cases in Figure  \ref{fig:tprofs_vs_mods_a}, results are generally similar, but extended deeper with the noticeably higher pressure of the thermal and visible photospheres (red and gray dotted lines, respectively). }
    \label{fig:tprofs_vs_mods_d}
\end{figure*}

\section{Cloud Optical Depth maps} 
\label{sec:appendix_cloud_maps}

\begin{figure*}[p]
    \centering
    \includegraphics[scale=0.58]{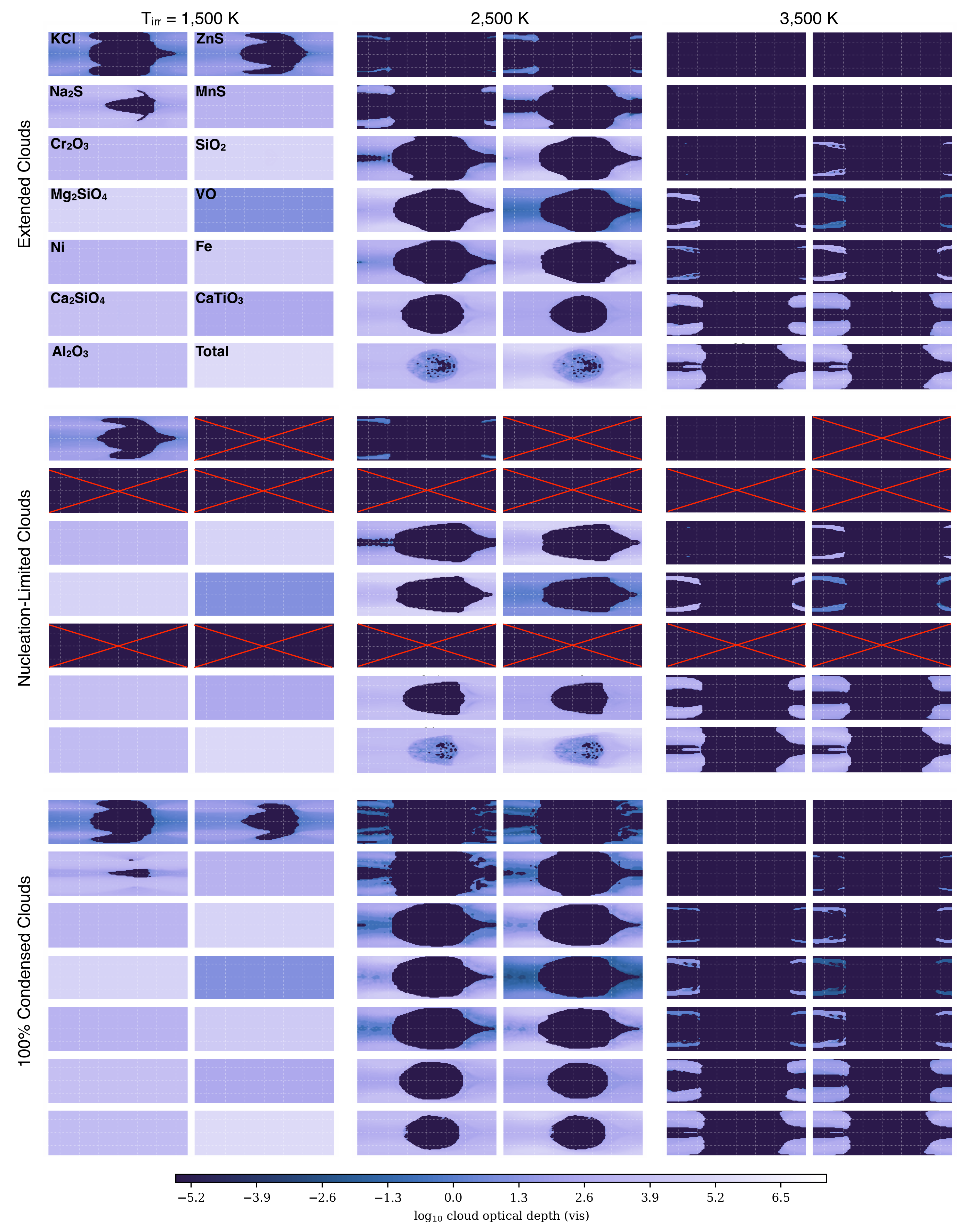}
    \caption{Species-by-species cloud maps for models with extended clouds, nucleation-limited clouds, and 100\% condensed (by mass) clouds at $T_{\mathrm{irr}}$ = 1,500, 2,500, and 3,500 K, as indicated.  Labeling is as in Figure~\ref{fig:cloud_maps}. Cloud optical thicknesses are measured from the top of the model down to the approximate pressure of the clear visible $\tau$=2/3 level (280 mbar with $g = 10$ m s$^{-2}$).}
    \label{fig:cloud_maps_extended}
\end{figure*}

\begin{figure*}[p]
    \centering
    \includegraphics[scale=0.58]{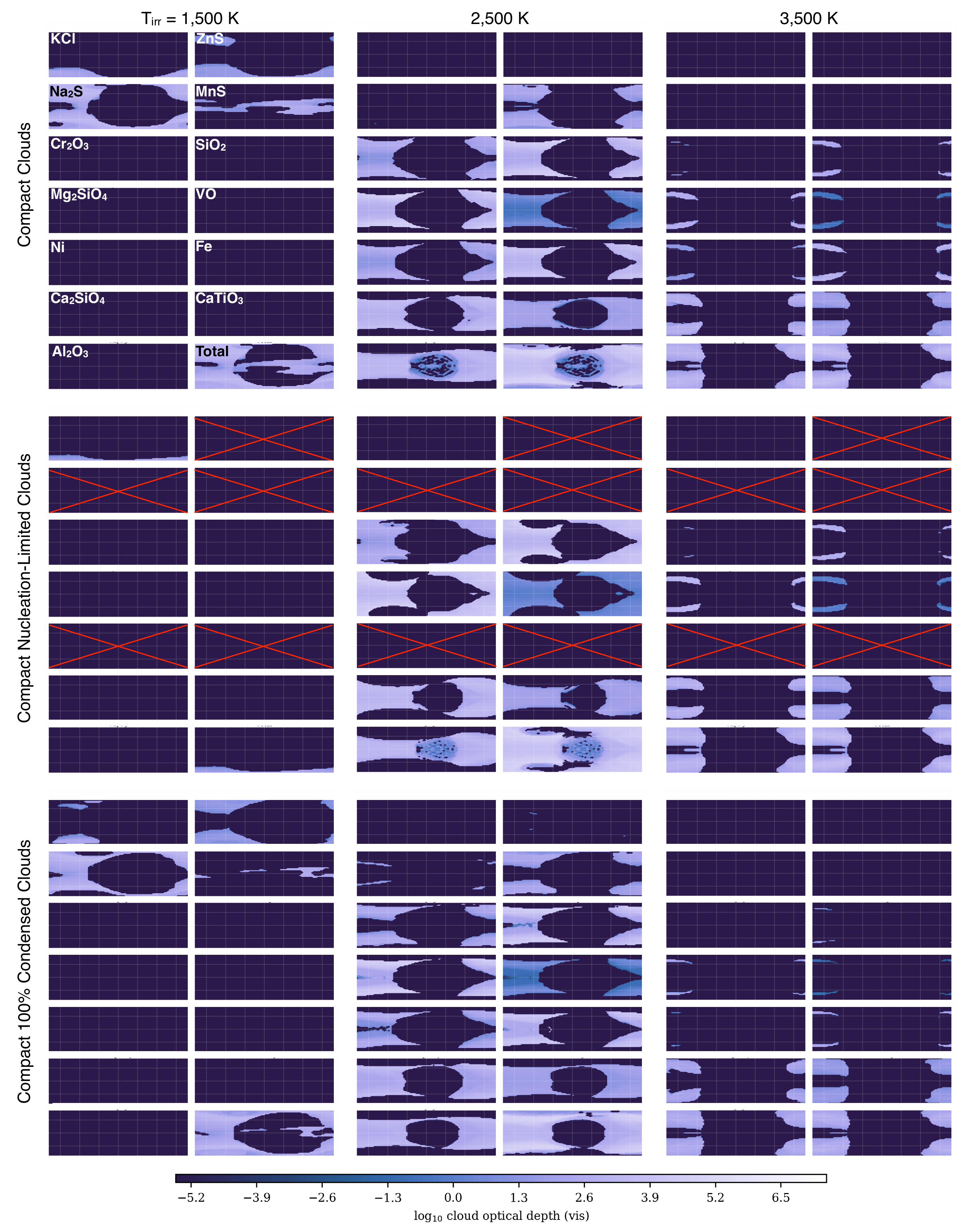}
    \caption{Species-by-species cloud maps for models with compact clouds, compact nucleation-limited clouds, and compact 100\% condensed (by mass) clouds at $T_{\mathrm{irr}}$ = 1,500, 2,500, and 3,500 K, as indicated.  Labeling is as in Figure~\ref{fig:cloud_maps}. Cloud optical thicknesses are measured from the top of the model down to the approximate pressure of the clear visible $\tau$=2/3 level (280 mbar with $g = 10$ m s$^{-2}$).}
    \label{fig:cloud_maps_compact}
\end{figure*}

The clouds in our models are made up of 13 individual condensable species. In Figures~\ref{fig:cloud_maps_extended} and \ref{fig:cloud_maps_compact} we show the species-by-species cloud maps for select models for three different irradiation temperatures spanning most of our model grid. These maps show the integrated visible-wavelength optical thickness of each individual cloud species above the height of the clear visible photospheric pressure (i.e.\ $\sim$280 mbar, where the visible optical depth equals $\sim$2/3 and the incident stellar radiation is reduced by roughly half). These maps are for cases with a surface gravity of 10 m s$^{-2}$, although maps with $g=40$ m s$^{-2}$ are very similar since the gas and cloud opacities both scale with gravity, as discussed in the main text. Maps of thermal optical thicknesses above the thermal photospheres show roughly similar distributions, but differ due to the lower pressure of the thermal clear-atmosphere photospheres ($\sim$60 mbar when $g=10$ m s$^{-2}$) and the significant wavelength-dependence of the cloud optical properties (see Appendix \ref{sec:appendix}).  The maps in Figures~\ref{fig:cloud_maps_extended} and \ref{fig:cloud_maps_compact} show how the contributions from the individual cloud species combine to produce the net trends in overall cloudiness seen in our results.

\end{document}